\documentclass{aa}

\usepackage{color}

\usepackage{graphicx}
\usepackage{txfonts}
\usepackage{natbib}

\newcommand{\oldbtxt}[1]{{#1}}

\newcommand{\jd}[1]{\,{\rm #1}}

\bibpunct{(}{)}{;}{a}{}{,} 
\begin{document}


\title{Chaotic and stochastic processes in the accretion flows of the black hole X-ray binaries revealed by recurrence analysis}
\author{Petra Sukov\'a\inst{1}, 
Mikolaj Grzedzielski\inst{1},
\and Agnieszka Janiuk\inst{1}
}

\institute{Center for Theoretical Physics,
Polish Academy of Sciences, Al. Lotnikow 32/46, 02-668 Warsaw, Poland \\
\email{psukova@cft.edu.pl}
}

\date{Received ...; accepted ...}
  
\abstract
{}
{The black hole candidates exhibit fast variability of their X-ray emission on a wide range of timescales. 
If the variability is recurrent, but not strictly periodic, we can find excess of power about certain frequency, often with Lorentzian shape. These peaks are referred to as quasi-periodic 
oscillations and they range between mHz up to kHz.
Their origin is still under debate and lots of different scenarios has been proposed, including resonant effects in the 
black hole accretion flow. 
The purely stochastic variability that 
occurs due to turbulent conditions in the plasma, is quantified by the power 
density spectra and appears practically in all types of sources and their spectral states. The specific kind of quasi-periodic flares is expected, when the
global structure of the accretion flow, governed by the nonlinear hydrodynamics, induces fluctuations around a fixed point solution. These conditions, which 
occur at high accretion rates, lead to the variability in the sense of 
deterministic chaos.  One of plausible mechanisms proposed to explain such variability is the intrinsic thermal-viscous instability in the accretion disk.
}
{We study the nonlinear behaviour of X-ray sources using the recurrence analysis method. We estimate quantitatively the indications for deterministic chaos, such as the R\'enyi's entropy, for the observed time series, and we compare them with the surrogate data.  This powerful method, widely known in other fields of physics, is used for the first time in the astrophysical context. 
}
{Using the data collected by \textit{RXTE} satellite, we reveal the oscillations pattern and the observable properties of six black hole systems. 
For five of them, we confirm the signatures of deterministic chaos being the driver of their observed variability.
}
{Both the well known microquasar GRS 1915+105, as well as its 
recently discovered analogue, IGR J17091-3624, exhibit variability characteristic to deterministic chaotic system. Therefore the underlying nature of the process must be intrinsically connected in these sources with the accretion flow instability, that leads to the limit cycle oscillations around a fixed point. 
\oldbtxt{Furthermore, we found significant traces of non-linear dynamics also in three other sources (GX 339-4, XTE J1550-564 and GRO J1655-40), particularly in the disk dominated soft state, as well as in the intermediate states at the rising and declining phase of the outburst. }
}
{}

\authorrunning{P. Sukov\'a et al.}
\titlerunning{Chaotic variability in BHXBs}

\maketitle

\section{Introduction}
\label{sect:intro}

The high energy radiation emitted by
black hole X-ray binaries originates in an accretion disk, where the viscous
dissipation of the gravitational potential energy is responsible for both
heating of the plasma and transporting of the angular momentum.
Hardly any of the observed black hole systems exhibits the radiation flux being
constant with time. Instead, most of the sources undergo fast and complicated
variability patterns on different timescales.
The variations that are purely stochastic in their nature,
are expected since the viscosity of the accretion disk is connected with
its turbulent behaviour induced by magnetic instabilities.
\oldbtxt{Furthermore, when the variations occur close to some frequency, 
the excess in the power spectrum known as the quasi-periodic oscillations (QPOs) appears.
The mechanism of the origin of QPOs is generally not known and different 
scenarios of their origin has been proposed, including oscillations of accretion disc/torus \citep{titarchuk2000global,2000ApJ...531L..41C}
or of shock fronts 
inside the accreting material \citep{2003ApJ...588L..89D,Sukova21022015} and
non-linear resonances between the radial and orbital epicyclic motions
 \citep{Abramowicz20061689}, possibly also relating the high frequency and low frequency QPOs 
(see the review by \citet{2014SSRv..183...43B}).
Among the non-resonant class of models, the Lense-Thirring precession is often invoked 
and aimed to explain the twin kHz QPOs as well as the low frequency QPOs, and was originally proposed for the
neutron star X-ray binaries in their Z and Atoll classes \citep{1998ApJ...492L..59S}.
}


In the models which invoke the global conditions in
 the accretion flow to be such that the system finds itself in an unstable
configuration, the large amplitude fluctuations around the fixed point
solution are induced. The variability of the disk that reflects its global evolution governed by
the nonlinear differential equations of hydrodynamics, will not be 
purely stochastic, but rather the observed behaviour of the disk will
be characterized by the deterministic chaos kind of process.\oldbtxt{This class of models which explains the quasi-periodic flares with the
global evolution of the accretion disk, rather than focusing on the flow oscillations
in some particular radii, is non-resonant at all. 
}

\oldbtxt{The flare-like events of the microquasars such as those observed in
 IGR 17091-3624 or GRS 1915+105 in their 'heartbeat' states 
are very strong, almost coherent low-frequency QPOs. 
In these sources, they are so orderly and quasi-coherent that it is possible to see the oscillations even directly in their 
light curves.
Generally, it is not possible to identify the QPOs by simply looking at their light curves. 
A usual method is to apply the Fourier transform to the light curve and to study the power spectrum 
in order to identify the QPO as a small peak, or a feature, just up the Poisson noise of the power spectrum.
Here we apply a novel mathematical method to study the quasi-periodicity and 
to determine the nature of the variability of the
X-ray sources, i.e., if it is stochastic or deterministic.
In the latter case, we propose, that the underlying mechanism of the deterministic chaotic behaviour
of the accretion flow is an underlying system of non-linear equations that governs its global evolution.
}

\subsection{Accretion disk instability due to the non-linear hydrodynamics}
The problem of chaotic variations in XRBs accretion disks is interesting
and still unsolved from both theoretical and observational point of view.
First, the classic theory of accretion as proposed by \cite{1973A&A....24..337S} and \cite{1974ApJ...187L...1L} predicts that the
disk is unstable and undergoes the limit cycle oscillations, if the
viscous stress tensor scales with the total (gas plus radiation) pressure, and 
the global accretion rate is large enough for the radiation pressure to dominate.

In general, the accretion disks may undergo the 
limit-cycle oscillations around a fixed point 
due to the two main types of thermal-viscous 
instabilities.
These are induced either by the domination of radiation pressure 
in the innermost parts of the disk, which occurs for high accretion rates, 
or by 
the partial ionization of hydrogen in the outer disk parts, for
 appropriate temperatures.
 Both these instability types are 
known for over 40 years in theoretical astrophysics. 

The hydrogen ionization instability operates in the outer regions of the 
accretion disk, where the temperatures are in the range 
of $\log T = 3.5-4$ [K] and the hydrogen is 
partially ionized. Under such conditions the opacities in the plasma
depend inversely on density and temperature, and non-linear processes
in the accretion flow are induced.
This type of instability leads to the outbursts of Dwarf Novae, 
and Soft X-ray transients, and generally its characteristic timescales
are on the order of months in case of stellar-mass accreting objects (for review see \cite{2001NewAR..45..449L}).

Another type of non-linear process, that operates on much shorter timescales,
is the instability induced by the dominant radiation pressure.
In classical theory of \citet{1973A&A....24..337S}, the accretion flow 
structure is based on
$\alpha$ prescription for the viscous energy dissipation.
It assumes that the non-zero component $T_{r \phi}$
of the stress tensor is proportional to the total pressure.
The latter includes the radiation pressure component which scales 
with temperature as $T^{4}$ and blows up in hot disks for large accretion rates.
This in turn affects the heating and cooling balance, between the energy 
dissipation and radiative losses.
If the accretion rate is small, then most of the disk is gas pressure dominated and stable. For large enough accretion rates, there appears a zone 
where some of its annuli are dominated by radiation pressure and unstable.
The larger the global accretion rate, the more annuli are
affected by the instability, starting at the inner edge which is the hottest.

If there was no stabilizing mechanism, the radiation pressure dominated disk would not survive. This is because in such a solution, the 
decreasing density leads to the runaway temperature growth. 
In consequence, the local accretion rate increases and more material is 
transported inwards. The disk annulus empties because of both increasing 
accretion rate and decreasing density, so there is no self-regulation 
of the disk structure.

However, the so called 'slim-disk' solution, where advection of energy
provides additional source of cooling acts as a stabilizing branch. 
In this way, advection allows the disk to survive and oscillate 
between the hot and cold states. Such oscillating behaviour leads to periodic 
changes of the disk luminosity. 

The theoretical calculations of the global, long-term disk evolution 
that are based on the 
vertically averaged structure, clearly show the oscillatory type of
behaviour, with characteristic limit-cycles \citep{2002ApJ...576..908J}.
The numerical computations performed in the 3-D shearing-box simulations 
with various codes either deny
\citep{2007ApJ...664.1045K}
or confirm \citep{2013ApJ...767..148J,2009ApJ...704..781H} that the MHD stress tensor scales with the total pressure. Therefore the issue of
$\alpha$-viscosity parametrization and contribution of the radiation pressure to
the energy dissipation is still an open problem.
Moreover, the 3-D simulations do not tackle the global structure of the flow, and therefore are not able to be verified directly with the observations of 
astrophysical sources.


\subsection{Observational support for the accretion instability}

For many years, the only source which undoubtfully showed oscillations that
are characteristic for the accretion disk instability due to the dominant radiation pressure, was the microquasar GRS 1915+105, in some of its observed states
\citep{2000A&A...355..271B,2002ApJ...576..908J}. 
The source was thought therefore to be 
unique of its kind, until the very recent discovery of another microquasar,
IGR J17091-3624, which was found to be an analogue of the
previously known microquasar \citep{2012MNRAS.422.3130C,2012ApJ...747L...4A}.
The recent hydrodynamical simulations of the global accretion disk 
evolution confirmed that the quasi-periodic flare-like events observed in 
IGR J17091, in some of its variability classes, are also
in a good quantitative agreement with the radiation pressure instability model
\citep{2015A&A...574A..92J}.

On the other hand, for other sources than these two microquasars,
no such detailed analysis was performed, neither observationally nor theoretically. However, as proposed by \citep{2011MNRAS.414.2186J}, 
at least eight of the known BH X-ray binaries should have their 
Eddington accretion rates large enough for the radiation pressure 
instability to develop. These
other black hole binaries can also be promising objects
for the radiation pressure instability, because they have their 
Eddington ratio above $\sim 0.15$. 
The particular value of this critical accretion rate depends on whether
the viscous torque parametrization is adopted in a modified version, such as
  $\alpha \sqrt{P_{\rm gas}P_{\rm tot}}$, instead of $\alpha P_{\rm tot}$.
Nevertheless, for the stellar mass black hole sources, the detailed analysis
and verification of available observations of various sources, 
should give constraints for
the presence of non-linear behaviour of their accretion flows.

Finally, because the instability timescale is scaling 
with the black hole mass, the intermittency in quasars on 
the timescales of hundreds to thousands of years is likely to be 
of a similar origin as in the microquasars. Here also some indirect 
arguments are proposed for the intermittent type of activity in the supermassive black holes, e.g. from the studies of the radio maps \citep{2010ApJ...718.1345K}. Also, the statistical studies occurred to be useful to provide an evidence for the episodic activity in AGN \citep{2009ApJ...698..840C}.

\subsection{Our current approach}

In the current work, we aim to tackle the problem of stochastic versus 
deterministic nature of the black hole accretion disk variability from the analytic and observational point of view.
We perform the recurrence analysis, which is a powerful tool used to study the time series, and known to work in broad range of applications, ranging from economy to geophysics and medicine \citep{0295-5075-4-9-004,Marwan2007237}. The recurrence analysis method for processing the time series was used recently also for studying chaos in motion of test particles \citep{kopavcek2010transition, sukova2011chaotic, sukova2012recurrence}.
\oldbtxt{Here, for the first time we show that this method has a great power in the studies of time series observed from astrophysical sources.

The essence of recurrence analysis method is to reveal the dynamical parameters of the system from the observed time series. These parameters, such as the \oldbtxt{R\'enyi's entropy (see equation \ref{K2_def} and the following text) or the maximal Lyapunov exponent, which characterizes the rate of stretching or contracting of the attractor \citep{ott2002chaos}}, may give indication that the underlying variability is deterministic in its nature. This is possible if the structure evolution equations of the dynamical system do not contain time explicitly, but are nonlinear and have unstable steady state solutions. The so-called 'recurrence plot' contains the information about time correlation and has a form of an array of points
in an $N x N$ square for a time series $u_{i}$, where $i=1...N$. The time series is used to construct the orbits $x(i)$ in the system's $d$-dimensional 
phase space (where $x$ may relate to the systems physical state, such as its density or temperature, and $d$ relates to the number of non-linear equations that govern its structure evolution).
The point in the recurrence plot is marked, whenever the trajectory returns close to itself, so that $x(j)$ is close to $x(i)$, i.e. closer
 than some assumed radius $\epsilon$ of an $d$-dimensional sphere. The plot is by definition symmetric with respect to the diagonal $i=j$, and the diagonal lines are not infinite, if only the variability is not completely periodic.  
}

We study the occurrence of long diagonal lines in the recurrence plots of observed data series and compare it to the series of surrogate data.
The latter are produced numerically
to have the same \oldbtxt{ mean and variance (shuffled surrogates, see below Sec. \ref{sect::General-approach}) or even the same power spectrum (IAAFT surrogates), albeit being created by random processes as the permutation of the original time series. 
The surrogate data were used as a counter-check with the correlation dimension technique and the single-value decomposition technique already by 
\citet{misra2006nonlinear}
in the context of the nonlinear variability in the microquasar GRS 1915+105.
Here also, our estimated dynamical invariants, e.g. the second order R\'enyi's entropy, which is a measure of the positive Lyapunov exponent and indication of deterministic chaos, are determined for both the observed series and the surrogates.
The example of recurrence plots of several observations and their surrogate is given in Fig.~\ref{fig:fig4}.}

We analyze the sample of BHXBs in order to determine the nature of their 
X-ray lightcurves. 
Our goal is to find the hints for deterministic chaotic behaviour
of the accreting black hole systems, as observed by the advanced X-ray satellites with good temporal resolution.
Thus we verify the results on the nonlinear behaviour of the GRS 1915+105
based on the RXTE data \citep{misra2006nonlinear} and we pursue
our analysis to other sources. In particular, we study IGR J17091-3624, in the four of its variability states.
We further study the sample of several less certain candidates for 
the globally unstable accretion disk evolution, and ultimately we try to answer
whether the two microquasars mentioned above are unique in their nature. 
The results of our analysis should allow for deeper insight into 
the accretion problem and to distinguish whether the variability is 
governed mainly by the set of nonlinear differential equations describing the 
global disk structure, and hence the fundamental physics, 
or rather by the local changes in the flow, and hence environmental 
conditions different for each source.

The article is organized as follows. In Section \ref{sect:observations}, we present our full observed sample of black hole X-ray binary sources and we briefly 
describe the RXTE data extraction procedure. 
In Section \ref{sect:analysis} we shortly describe the recurrence analysis method and its general properties, while the details about our method are given in the Appendix~\ref{sect:Recurrence}. We summarize the main results for the microquasar IGR J17091-3624 in Section~\ref{sect:IGR_sum} and in App.~\ref{Sect:Comparison} we show in details how the method was applied to this source. In App.~\ref{sect:poincare} we give some more details about the significance of the method, depending on the strength of the noise which is always present in the observed data sets.
In Section \ref{sect:other_sources} we apply the recurrence analysis method further to the rest of our observed sample of the black hole X-ray binaries. 
Finally, in Section \ref{sect:diss} we discuss our results and we present conclusions.

\section{Observations}
\label{sect:observations}

We analysed several tens of observations of six black hole X-ray binaries, in which the radiation pressure instability is considered to develop \citep{2011MNRAS.414.2186J,2015A&A...574A..92J}. We expect these sources lightcurves to exhibit signatures of non-linear behaviour. Below we describe briefly the sources that belong to our sample and the method which we used for the extraction of the RXTE archival data.

\subsection{Our sample}
\label{sect:sample}

\begin{table}
\begin{center}
\begin{tabular}{|c|c|c|c|}
\hline
 & & &number\\
ObsID&date&class&of points\\
\hline \hline
01-00&2.3.2011&HIMS&3380\\ \hline
01-02&3.3.2011&SIMS&5536\\ \hline
&&SIMS/&\\
02-00&4.3.2011&HIMS&3174\\ \hline
02-02&8.3.2011&HIMS&6383\\ \hline
02-03&10.3.2011&HIMS&3305\\ \hline
03-00&12.3.2011&SIMS&3030\\ \hline
03-01&14.3.2011&IVS&6682\\ \hline
04-00&19.3.2011&$\rho$&5472\\ \hline
04-01&23.3.2011&IVS&2297\\ \hline
04-02&22.3.2011&$\rho$&5939\\ \hline
04-03&24.3.2011&IVS&5499\\ \hline
05-00&27.3.2011&$\rho$&5444\\ \hline
05-01&30.3.2011&$\rho$&2872\\ \hline
05-02&25.3.2011&SIMS&6620\\ \hline
05-03&29.3.2011&$\rho$&4683\\ \hline
05-04&31.3.2011&$\rho$&5684\\ \hline
06-00&2.4.2011&$\rho$&5227\\ \hline
06-01&3.4.2011&$\rho$&6408\\ \hline
06-02&5.4.2011&$\rho$&5611\\ \hline
06-03&6.4.2011&$\rho$&3617\\ \hline
07-00&10.4.2011&$\rho$&3201\\ \hline
07-01&11.4.2011&$\rho$&6389\\ \hline
07-02&12.4.2011&$\rho$&6388\\ \hline
08-00&15.4.2011&$\rho$&6365\\ \hline
\end{tabular}
\end{center}
\caption{ Observations of IGR J17091-3624. The prefix of the ObsID is 96420-01-. In the third column the spectral class according to \cite{0004-637X-783-2-141} is provided.  The number of points of the observation used for the analysis in in the last column, the time bin is 0.5s. The data were extracted from the generic event mode data using events from PCU2, channels 5-24. \label{Table1}}
\end{table}

 The studied sources are the microquasars IGR J17091-3624, GRS 1915+105, 
GRO 1655-40, XTE J1650-500,  XTE J1500-564 and GX 339-4. \oldbtxt{Apart from the first one, which went into an outburst later in 2011,} they were selected \oldbtxt{from the list given in \citep{2011MNRAS.414.2186J}} based on the presumably high accretion rate to the Eddington rate ratio and \oldbtxt{high count rate at the same time.  We included also the source GX 339-4, because higher accretion rate was reported by \cite{Zdziarski01072004} combined with the results of \cite{2004ApJ...609..317H} than which was assumed in \cite{2011MNRAS.414.2186J} and because this source shows also other interesting features such as low frequency QPOs with changing frequency.}

The first two systems are known to exhibit flare-like events in some states \oldbtxt{(see the references afterwards)} and their \oldbtxt{similar} non-linear behaviour has been studied by different methods \oldbtxt{\citep{misra2004,misra2006nonlinear,2041-8205-742-2-L17,Pahari11122013,0004-637X-778-1-46}}. We decided to built-up our method on the sample of IGR J17091-3624, whose spectral states have been studied by \cite{0004-637X-783-2-141}. \oldbtxt{The light curve of IGR J17091-3624 during the heartbeat state has been modelled by \cite{2015A&A...574A..92J} with the model of the disc which undergoes the radiation pressure instability induced oscillations.
The wind ejected from the accretion disk regulates the amplitude of the oscillations and increase its frequency, so that the simulations are in agreement with the observations. Moreover, the stronger winds, which are detected in some of the non-heartbeat states via the X-ray spectroscopy, suppress the oscillations completely. Therefore, the idea of the non-linear instability producing the flares in IGR J17091-3624 is strongly supported by this modeling.}

\oldbtxt{Further, we } confirm the results of \cite{misra2004} for some of the observations of GRS 1915+105 \oldbtxt{by our method}.

Later we applied our method for several observations of the other four sources.  We have not carried out an extended study of these sources, we rather aimed to find several examples of observations, which do or do not show the traces of the non-linear behaviour. We took into account the observations with high enough count rate and we focused on the rising and declining phases of the outbursts. At that time the given source goes through the spectral state transition, which can be correlated with an unstable stage of the accretion disc prone to the internal instability. 

Even though our sample of these four objects is limited and serves as a brief example rather than a deep survey, we have found indications of non-linearity in several cases, apart from the two aforementioned microquasars.

\begin{figure}
\includegraphics[width=\columnwidth]{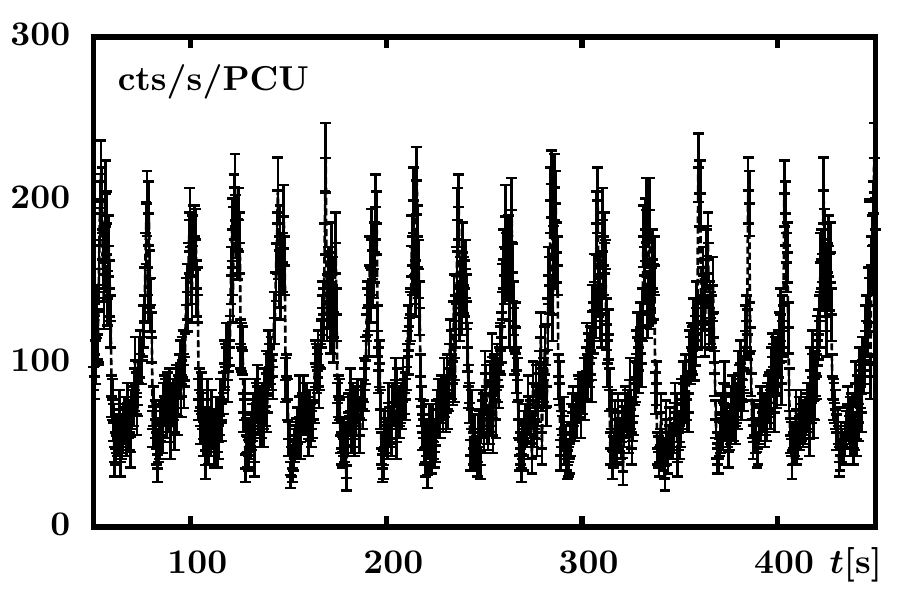}
\caption{ Lightcurve of the microquasar IGR J17091-3624, for the observation ID 96420-01-06-02, extracted in the energy range 2-10 keV with time bin ${\rm d}t=0.5$s. We computed the significance of the non-linear dynamics detected in this observation $\bar{\mathcal{S}}=3.72$.}
\label{fig:602_lc_IGR}
\end{figure}


\subsubsection{IGR J17091-3624} \label{sample:IGR}

The microquasar IGR J17091-3624 is a moderately bright transient X-ray binary 
(peak flux level at $\sim$20 mCrab in the range 20-100 keV) discovered
in 2003 \citep{Kuulkers}.
The 2011 outburst was the brightest one ever observed from IGR J17091, and the source flux increased up to 120 mCrab in the range 2-10 keV ~\citep{2012MNRAS.422.3130C}.
The \textit{RXTE}/PCA data showed quasi-periodic flare-like events occurring at a rate
between 25 and 30 mHz~\citep{Altamirano11a}, which resembled the
`heartbeat' variation observed previously in the BH binary GRS 1915+105.

We tested our method on the sample of observations of IGR J17091-3624 selected 
from those analysed in \cite{0004-637X-783-2-141}, which \oldbtxt{ according to their classification}  belong to different 
spectral classes (hard intermediate state (HIMS), soft intermediate state 
(SIMS), intermediate variable state (IVS) and variable state (also denoted as 
$\rho$ state)). All the observations belong into the proposal number 96420, 
target 01 and they were taken by RTXE satellite between March and April 2011.
The studied observations of IGR J17091-3624 are summarized in Table~\ref{Table1}
and an exemplary lightcurve of the source is shown in Figure \ref{fig:602_lc_IGR}.


\subsubsection{GRS 1915+105}

\begin{figure}
\includegraphics[width=\columnwidth]{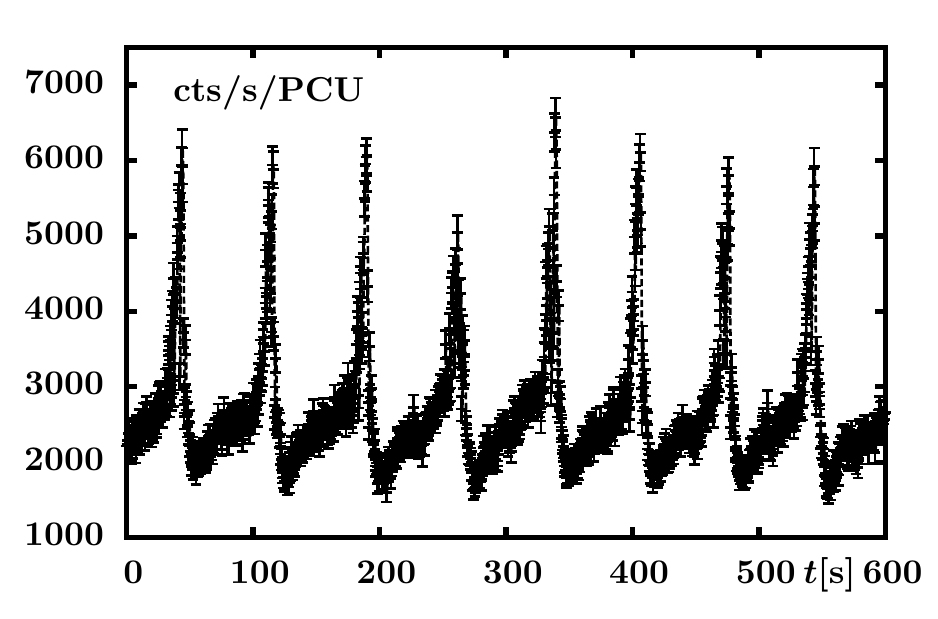}
\caption{Lightcurve of the microquasar GRS 1915+105, for the observation ID 20402-01-03-00, extracted in the full energy range 2-60 keV with time bin ${\rm d}t=0.5$s from Standard 1 data in the $\rho$ state (classification according to \citealt{misra2004}).}
\label{fig:602_lc_GRS}
\end{figure}

GRS 1915+105 is \oldbtxt{ a very well known} low mass X-ray binary, discovered in 1992 
\citep{1992IAUC.5590....2C}. 
Its behaviour was interpreted with the time-dependent evolution of an accretion disk, which are thermally and viscously unstable \citep{1997ApJ...485L..83T, 2000ApJ...542L..33J}.
~\citet{2000A&A...355..271B} classified for GRS 1915+105 the 14 classes of variability, out of which some are so called 'heartbeat states', and exhibit regular amplitude periodic oscillations, like the one shown in the lightcurve on Figure \ref{fig:602_lc_GRS}.
The flux emission during the heartbeat states of the source reaches the Eddington luminosity ~\citep{2004MNRAS.349..393D}. 
The average mass accretion rate in GRS 1915+105 
on the order of $10^{18} - 3 \cdot 10^{19} $g s$^{-1}$
(0.031- 0.63 in Eddington units) was estimated by \cite{2011ApJ...737...69N}.



\subsubsection{GX 339-4}

X-ray nova GX 339-4, discovered in year 1972 
\citep{1997AstL...23..433L}, exhibits high level of X-ray activity.

The distance was estimated as $\gtrsim 7$ kpc by \cite{Zdziarski01072004}, who also reported that the luminosity varied during its outbursts (15 outburst from 1987 to 2004) between $0.01 - 0.25 L_{\rm Edd} (d/8 {\rm kpc})^2 (10M_\odot/M)$. Similar values were also confirmed by \citep{2011MNRAS.416..311K}. However, \cite{2004ApJ...609..317H} proposed the possibility of the location on the far side of the Galaxy with distance $d \gtrsim 15$ kpc, which would imply that the source reaches the Eddington luminosity in the peaks.


\subsubsection{GRO J1655-40}

Low mass X-Ray binary GRO J1655-40 was discovered on Jul, 27, 1994 with BATSE during the outburst and soon after also its optical eclipsing counterpart has been found. 
\cite{1997ApJ...477..876O} found the mass of the black hole $M=7.02\pm0.22M_\odot$
and estimated the averaged mass transfer rate as $3.4 \cdot 10^{-9} M_\odot {\rm yr}^{-1} = 2.16 \cdot 10^{17} {\rm g\, s}^{-1}$.

A strong highly ionized accretion disc wind is reported by \cite{0004-637X-680-2-1359} on Apr, 1, during the 2005 outburst. The origin of the wind is likely not driven by the thermal process. 


\subsubsection{XTE J1550-564}

XTE J1550-564 was first discovered
by the ASM on board RXTE on 1998 September 7 during outburst. 


During the 1998 outburst the source exhibited a low and intermediate frequency QPOs with evolving frequency in the range 0.08 - 13 Hz \citep{Chakrabarti11042009} \oldbtxt{ and the accretion rate was a significant fraction of Eddington luminosity \citep{2010MNRAS.403...61D,2011MNRAS.411L..66H}. }


\subsubsection{XTE J1650-500}
XTE J1650-500 is the black hole X-Ray binary with black hole mass $4 - 7.3 M_\odot$ \citep{2004ApJ...616..376O,2008A&A...492..319S}. 
The object emits flares with non-thermal energy spectra and occur when the persistent luminosity is near $3 \cdot 10^{34} $ erg s$^{-1}$(d/4 kpc)$^2$  in 2.8-20 keV range \citep{2003ApJ...592.1100T}.
\cite{2006MNRAS.366..235H} estimated the distance based on the luminosity during the spectral state transition as $2.6\pm0.7$ kpc assuming the black hole mass $M=4M_\odot$, for which the Eddington luminosity $L_{\rm Edd}\sim1.1\cdot10^{37}$ erg s$^{-1}$ in 0.5 keV - 10 MeV.

\subsection{Data analysis}
\label{sect:data}

We analyzed the online data from the RXTE satellite.
We extracted the time series 
using \texttt{Heasoft 6.16} high energy astrophysics software package. 
 Because we want to compare our results for IGR J17091 with the classification of \cite{0004-637X-783-2-141}, we use the filtering mentioned in that paper using data from all xenon layers from the PCU counter number 2 and we processed the events from channels 5 to 24 (2-10 keV approximately).
Afterwards we selected several observations for the other 5 sources, which were observed by RXTE between 1996 and 2010. These observations were made in different observational modes with different data available. For each source we selected data in lower energy band or in full energy range and normalized it to count/s/PCU. These sources and details about the data extraction are summarized in Table~\ref{Table2} in Section \ref{sect:other_sources}, together with the most important results of our 
theoretical analysis.


 We adjust the proper binning 
 time to minimize the error and simultaneously to not lose the information about oscillations at the scale of several seconds.
We use the minimal binning time between $0.032$s and $0.5$s. Exact value depends 
on the flux and on the time scale of the variability in the data. Therefore, 
 for the flux on the order of 
hundred counts per second we use the minimal binning size $0.5$, if the average flux per PCU exceeds $500$ cts/s. For $1000$ cts/s, we can drop the value to 
$0.2$s, $0.1$s or even $0.032$ s. 
However, in some cases the variability is slower and we can increase the time bin above the minimal level.

While extracting the data, we use the \textit{xdf} tool, and  filter the data with \textit{xtefilt}, and we make the 'good time interval'. For the science array format data files (standard-1 mode or generic single bit mode) we extract the data using \textit{saextrct} command, with a proper 
bin size. The standard-1 mode data are collected in the full energy spectrum of PCA (2-60 keV), while the single bit mode contains one channel band in the lower energy range (it usually contains channels 0-35).
For the science event format data files (generic event data mode), we extracted the lightcurves with the \textit{seextrct} command with the energy range corresponding approximately to 2-10 keV.
The resulting count rate is normalized to number of PCUs, which were used for the data extraction. The data mode used for the extraction is indicated in Table~\ref{Table2} in Section \ref{sect:other_sources}.

The background subtraction is ignored in our analysis. This is because for the 
bright sources, it can be safely neglected, and also due to the fact that in RXTE observations the background is not measured, but simulated numerically. 
We do not have therefore a background measurement in the function of time, while the essential part of our analysis is to catch up the changes of about 0.5 s time-scales.
 
\oldbtxt{
We have not done the spectral analysis of the observations. 
The spectra were studied and in some cases classified by other authors. 
For the spectral state we provide in table~\ref{Table2} their results or our guess based on the published plots (e.g. on the hardness ratio plot from \cite{0004-637X-680-2-1359} for the 2005 outburst of GRO J1655-40). 
}
 
\section{Analysis of the chaotic processes}
\label{sect:analysis}

Our aim is to reveal the important information about the BHXBs on the basis of their X-ray light curves \oldbtxt{ using the recurrence analysis.   }

The recurrence analysis is a well-established method for studying the properties of the dynamical systems based on the behaviour of its phase space trajectories. More specifically, it looks on the ``recurrences'' in the phase space, which are the moments, when the phase space trajectory returns close to itself, so that the points $\vec{x}(t_i)$ and $ \vec{x}(t_j)$ are close.  
The long diagonal lines (of length $l$) are formed in the recurrence plot representing the recurrence matrix (see relation \ref{RP_def}), when the pairs of successive points on the trajectory are close ($\parallel \vec{x}(t_i) - \vec{x}(t_j) \parallel < \epsilon, \parallel \vec{x}(t_{i+1}) - \vec{x}(t_{j+1}) \parallel < \epsilon, \dots, \parallel \vec{x}(t_{i+l-1}) - \vec{x}(t_{j+l-1}) \parallel < \epsilon $). These lines correspond to the case, that the later part of the trajectory evolves inside a tube of $\epsilon$ diameter along the previous piece. 

In case of a periodic orbit, the recurrence plot would consist of infinite diagonals spaced by the period~$T$. Stochastic processes yield randomly distributed points, which do not tend to form any lines or structures. 
Chaotic process produces lines with finite length, which is related to the \oldbtxt{maximal} Lyapunov exponent, because for chaotic systems nearby trajectories 
eventually diverge outside the $\epsilon$-tube.
Therefore, one of the basic but important quantifiers of the recurrence analysis is the length of the longest diagonal line $L_{\rm max}$.
We quantify it simply as the number of points composing the line.

The recurrence analysis can be supplied with the time delay reconstruction of the phase space trajectory \citep{takens1981detecting}, so that the measured time series of one dynamical quantity could be used for studying the underlying dynamical system. From that point of view the method is suitable for the analysis of the astrophysical data obtained e.g. by X-ray satellites, for which only the
lightcurve is available. 
Therefore, the motivation for us is also to explore the abilities of recurrence analysis in the context of variability of X-ray binaries.

Here, the underlying dynamical system under study is the accreting gas governed by a global physical parameter, 
e.g. by the mass accretion rate. 
If the accretion is stationary, the produced flux has constant mean and the variability is due to stochastic fluctuations. 
But if the global parameters are such that the intrinsic instability of an accretion disk develops, the mass accretion rate locally varies in time, which leads to deterministic changes of the flux, possibly also overlaid with some stochastic fluctuations. 

Detailed description of the main quantities of the recurrence analysis used in this paper is given in Appendix \ref{sect:Recurrence}. We show there several examples of the technique. We also encourage the interested reader to look further into the extensive study of the method given in \cite{Marwan2007237} and references therein. The details of the method, definitions of the terms and quantities together with discussion about the possible drawbacks and difficulties arising from the presence of noise are given in Appendices \ref{Sect:Comparison} and \ref{sect:poincare}.

\subsection{General approach}
\label{sect::General-approach}

The recurrence analysis 
technique needs careful setting of the parameters. For noisy short data series such estimation is always complicated.
Therefore instead of the direct estimate of these parameters, 
we rather incorporate the method of surrogate data, as proposed by \cite{Theiler199277}. 

We first pose the ``null hypothesis'' about the measured time series, e.g. that the data are product of temporally independent white noise or linearly autocorrelated Gaussian noise. 
Then we produce the set of surrogate series, which fulfill this hypothesis, but contain as much stochasticity as is possible (at given level of the hypothesis), hence there is no further hidden non-linear dynamics. 

In the case of the temporally independent white noise, the surrogates are created as the random \oldbtxt{ permutation} of the values of the time series, thus such surrogates (we call them ``shuffled surrogates'') have the same mean and variance as the original time series.
In the case of the linearly autocorrelated Gaussian noise, the surrogate series are constructed in such a manner, that they have the same distribution of values (flux per second per PCU for the observed lightcurves) and almost the same spectrum \oldbtxt{\citep{Theiler199277}}. 
This can be achieved by an iterative algorithm called Iterative Amplitude Adjusted Fourier Transform Algorithm (IAAFT) or its stochastic version SIAAFT, \oldbtxt{ which is described e.g. by \cite{Schreiber2000346,venema2006stochastic}. The algorithm iteratively replaces the amplitudes and magnitudes of the Fourier coefficients of the surrogate series (starting from the white noise) by corresponding values obtained from the original time series, so that at the end the surrogate series is a permutation of the original time series with almost the same spectrum.} Throughout this paper, we will refer to data series created by this procedure as simply ``surrogates'', or IAAFT surrogates.

For one experimental data series we typically construct one hundred of its surrogates. 
Afterwards, we apply the recurrence analysis both on the real experimental lightcurves and on the corresponding surrogates and we compare the obtained properties for the real and artificial data.

If the quantity measured for the real data differs significantly from the value obtained for its surrogate, we can reject our initial null hypothesis. It means that 
the non-linear behaviour and perhaps chaotic nature of the system appears in the case of IAAFT surrogates, or non-stochastic but linear behaviour in the case of shuffled surrogates. 
Luckily, there is actually no need for the discriminating criterion to be a particular physical quantity, hence the majority of problems with accurate choice of the parameters for the analysis are overcome by our approach.
The discriminating criteria are provided by the recurrence analysis of the time series.

\subsection{Numerical method}
\label{sect:recipe}

For some tasks, including the creation of the IAAFT surrogates, we employ several procedures from the publicly available software package {\tt TISEAN} \footnote{\tt http://www.mpipks-dresden.mpg.de/\textasciitilde tisean/ } \citep{1999chao.dyn.10005H,Schreiber2000346}. The shuffled surrogates are produced by our own code written in IDL. The recurrence analysis is performed on the basis of the software package described by \cite{Marwan2007237,RP-web}, which yields also the cumulative histogram of diagonal lines. The linear regression for $K_2$ estimation and other post-procession is done by our IDL codes.

Below, we summarize the main steps of the handling with the data.

\begin{itemize}
\item For every observation we extract the lightcurve (discussion about the time and energy bin is given in Sect.~\ref{sect:data} and in Table~\ref{Table2}).
\item We rescale the extracted light curve to have zero mean and unit variance before producing the surrogates for the ease of comparison between different observations using procedure {\tt rescale} from {\tt TISEAN}.
\item We use the procedure {\tt mutual} from {\tt TISEAN} in order to find appropriate guess of the time delay $\Delta t = k {\rm d} t$ (see App.~\ref{sect:Recurrence} and Fig.~\ref{fig:fig1}).
\item We use the procedure {\tt false\_nearest} from {\tt TISEAN} in order to find appropriate guess of the embedding dimension $m$ (see App.~\ref{sect:Recurrence} and Fig.~\ref{fig:fig2}).
\item We produce 100 surrogates (shuffled by our own code, IAAFT by the procedure {\tt surrogates} from {\tt TISEAN} -- or both) for each observation.
\item We find the recurrence threshold $\epsilon_{\rm min}$ and $\epsilon_{\rm max}$ such, that the recurrence rate (RR) of the observation is approx. 1\% and 25\%, respectively\footnote{ The recurrence rate is the ratio of the number of recurrence points to all points of the recurrence matrix.}. We create a sequence of $\epsilon$ in the range ($\epsilon_{\rm min},\epsilon_{\rm max}$) with constant difference. The number of used thresholds ranges between 10 to 40 depending on the length of the data series (and thus on CPU time demand of the analysis).
\item We construct the file with RQA quantifiers and the cumulative histograms of diagonal lines, by the program {\tt rp} described by \cite{Marwan2007237, RP-web} with the found parameters $k,m$ for all $\epsilon$'s for the observation and the surrogates.
\item For each $\epsilon$ we compute the estimate of the R\' enyi's entropy $K_2^{\rm obs}$ for the observation and $\{K_{2}^{\rm surr}\}_{i=1}^{100}$ for the set of surrogates. The R\'enyi's entropy of the order $\alpha$ is in general defined as follows:
\begin{equation}
K_{\alpha} = {1 \over 1-\alpha} \ln \sum_{i=1,N} p_{i}^{\alpha} 
\end{equation}
where $p_{i}$ is the probability that the random variable has a value of $i$, and basically describes the randomness of the system. For detailed explanation of the meaning of entropy $K_2$, see relations \ref{K2_def} and \ref{cumul} in Appendix \ref{sect:Recurrence}.
\item We compute the average significance $\bar{\mathcal{S}}_{\rm shf}$ with respect to the shuffled surrogates and the averaged significance $\bar{\mathcal{S}}$ with respect to the IAAFT surrogates.
The definition of significance of chaotic process is taken as:
\begin{equation}
\mathcal{S}(\epsilon) = \frac{N_{\rm sl}}{N^{\rm surr}} \mathcal{S}_{\rm sl} - {\rm sign}( Q^{\rm obs} (\epsilon) - \bar{Q}^{\rm surr}(\epsilon) ) \frac{N_{\mathcal{S}_K}}{N^{\rm surr}}  \mathcal{S}_{K_2}(\epsilon) , \label{significance}
\end{equation}
where $N_{\rm sl}$ is the number of surrogates, which have only short diagonal lines, and $N^{\rm surr}$ is the total number of surrogates, $Q^{\rm obs}$ and ${Q}^{\rm surr}$ are the natural logarithms of $K_2$ entropy for the observed and surrogate data, respectively, $\mathcal{S}_{\rm sl} =3$ and  $\mathcal{S}_{K_2}$ is the significance computed only from the surrogates, which have enough long lines according to the relation
\begin{equation}
\mathcal{S}_{K_2} (\epsilon) = \frac{| Q^{\rm obs} (\epsilon) - \bar{Q}^{\rm surr}(\epsilon) |}{\sigma_{Q^{\rm surr}(\epsilon)}}.
\end{equation}
The significance defined in this way expresses how much the value $K_2^{\rm obs}$ differs from the mean value $\bar{K_2}^{\rm surr}$ measured in the units of the standard deviation of the set $\{K_2^{\rm surr}\}_{i=1}^{100}$ in the logarithmic scale $\sigma_{Q^{\rm surr}(\epsilon)}$.
For further details, the reader is asked to read the explanation in App.~\ref{Sect:Comparison} and the discussion before equation \ref{significance}.  

\end{itemize}

We distinguish between fully stochastic behaviour (those observations, which show no significant difference compared to the shuffled surrogates, $\bar{\mathcal{S}}_{\rm shf}<1.5$ or those, which do not have enough long lines in the RP), non-stochastic, but linear behaviour (observations with a significant result with shuffled surrogates, but non-significant result with IAAFT surrogates, $\bar{\mathcal{S}}_{\rm shf}>1.5, \bar{\mathcal{S}} < 1.5$) and non-linear, possibly chaotic behaviour (observations with significant result with respect to the IAAFT surrogates, $\bar{\mathcal{S}} > 1.5$).

In Tables \ref{TableIGR_result} and \ref{Table2}  we provide the number of points of the observed time series, which is used for the analysis, the length of the longest line for several recurrence rates (the longer the diagonal lines are compared to the time delay $k$, the more regularly the system behaves), the averaged significance $\bar{\mathcal{S}}_{\rm shf}$ and/or $\bar{\mathcal{S}}$ and the number of thresholds for which the average is computed.

\section{Case study of IGR J17091-3624}
\label{sect:IGR_sum}

We used the sample of observation of IGR J17091-3624 as described in Section \ref{sample:IGR} and in Table~\ref{Table1} for testing the capabilities of recurrence analysis to reveal important information about the dynamical properties of the source.\oldbtxt{Details of the procedure are given in App.~\ref{sect:Recurrence} and \ref{Sect:Comparison}, main results are described here and summarized in Table~\ref{TableIGR_result}. }

\begin{table}
\begin{center}
\begin{tabular}{|c|c|c|c|c|c|c|}
\hline
 & &{\small 5\%}&{\small 10\%}&{\small 20\%}&&\\
ObsID&N&$L_{\rm max}$&$L_{\rm max}$&$L_{\rm max}$& $\bar{\mathcal{S}}$ & $N_\epsilon$\\
\hline \hline
01-00&3380&6&7&10&-&-\\ \hline
01-02&5536&6&7&11&-&-\\ \hline
02-00&3174&4&6&10&-&-\\ \hline
02-02&6383&6&7&10&-&-\\ \hline
02-03&3305&4&6&8&-&-\\ \hline
03-00&3030&4&6&8&-&-\\ \hline
03-01&6682&6&7&12&-&-\\ \hline
04-00&5472&73&144&198&1.93&32\\ \hline
04-01&2297&6&8&13&-&-\\ \hline
04-02&5939&33&117&274&3.21&27\\ \hline
04-03&5499&6&10&14&-&-\\ \hline
05-00&5444&12&22&100&2.35&18\\ \hline
05-01&2872&10&26&89&1.08&15\\ \hline
05-02&6620&6&6&10&-&-\\ \hline
05-03&4683&13&72&126&2.45&19\\ \hline
05-04&5684&13&68&164&3.18&19\\ \hline
06-00&5227&43&130&215&2.73&28\\ \hline
06-01&6408&125&280&344&3.29&35\\ \hline
06-02&5611&134&207&391&3.72&35\\ \hline
06-03&3617&19&77&79&1.21&18\\ \hline
07-00&3201&137&223&290&2.26&34\\ \hline
07-01&6389&125&171&289&3.04&34\\ \hline
07-02&6388&106&180&285&3.13&36\\ \hline
08-00&6365&35&51&100&1.63&17\\ \hline
\end{tabular}
\end{center}
\caption{ Results of our analysis for IGR J17091-3624. The prefix of the ObsID is 96420-01-. We used the RP parameters $m=10,\Delta t=7$s for every observation. The number of points $N$ of the observation used for the analysis is in the second column, the time bin is 0.5s. Next, we present the number of points composing the longest diagonal line for three different thresholds yielding the recurrence rate approx. 5\% (third column), 10\% (fourth column) and 20\% (fifth column).  The average significance $\bar{\mathcal{S}}$ and the number of different thresholds $N_\epsilon$ used for the averaging is in the two last columns. \label{TableIGR_result}}
\end{table}



We included in our study several observations from the hard intermediate state (HIMS), soft intermediate state (SIMS), intermediate variable state (IVS) and variable state (also denoted as $\rho$ state) of the source (see  \cite{0004-637X-783-2-141}).
All observations classified by \cite{0004-637X-783-2-141} as SIMS and HIMS and the observation 03-01 classified as IVS \oldbtxt{have low and almost constant values of mutual information (see Fig.~\ref{fig:fig1}).} 
Three observations (04-01, 04-03 (IVS) and 06-03 ($\rho$)) show smooth decrease but no second maximum.
All other $\rho$ observations \oldbtxt{show oscillations with period ranging from $10\jd{s}$ to $20\jd{s}$. }

We found that observations \oldbtxt{with no oscillations of mutual information}
 show no lines long enough for computation of $K_2$ (the criteria for the length and number of lines are described in App.~\ref{Sect:Comparison}) \oldbtxt{and their $L_{\rm max}$ behaves similarly like that of both types of surrogates.}. Therefore, these observations are consistent with the hypothesis of temporally independent white noise (see Fig.~\ref{fig:502Lmax}) and the source did not show any dynamical evolution during that time. 

The observations 06-03 and 05-01 do contain enough long lines, so that the significance can be computed. In both cases non-significant results are obtained with respect to the IAAFT surrogates.  
However, we obtain\oldbtxt{significance with respect to shuffled surrogates} $\bar{\mathcal{S}}_{\rm shf}=3$ for both,\oldbtxt{computing the} average using $N_\epsilon=13$ and $N_\epsilon=15$ different thresholds for 06-03 and 05-01, respectively. 
Thus, we classify these two observations as a non-stochastic linear process. 
As mentioned in Appendix \ref{sect:poincare}, this can also mean, that these observations refer to the period, when the source was following regular or ``sticky trajectory'' of the generally non-linear dynamical system. Such orbit for very long time evolves very close to a regular island, sharing its main dynamical properties, but eventually departures to another part of the phase space, differing significantly from the regular motion (for more details see e.g. \cite{semerak2012free}). \cite{0004-637X-783-2-141} classified these observations as the variable $\rho$ state.

The rest of the $\rho$ state observations show significant result with  $\bar{\mathcal{S}}>1.5$. The values range between 1.63 for ObsID 08-00 to 3.72 for ObsID 06-02. 
For these observations we claim that the source was showing non-linear behaviour.

\oldbtxt{ For illustration we also provide the information about the length of the longest diagonal line for different values of RR in Table \ref{TableIGR_result}. Higher ratio $L_{\rm max}/k$ means, that the system evolves similarly for longer time \oldbtxt{compared to the period of the oscillations (i.e. that the phase trajectory in two different times evolves within $\epsilon$-tube)}. 
Further details are given in App.~\ref{Sect:Comparison}.
}

\begin{figure}
\includegraphics[width=\columnwidth]{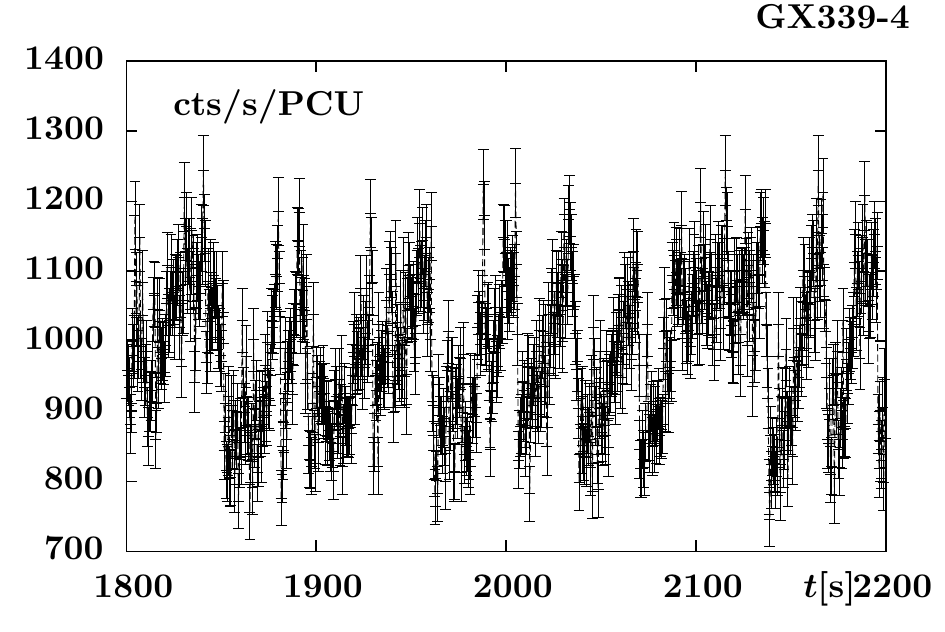}
\caption{Lightcurve of the source GX 339-4. The observation ID is 95409-01-16-05 and the source was in its soft-intermediate state.
Data were extracted in the energy range 2-10 keV (channels 5-25). 
We computed the significance of the non-linear dynamics detected in this observation $\bar{\mathcal{S}}=2.90$.}
\label{fig:GX339_lc}
\end{figure}

\section{Using the method on other sources}
\label{sect:other_sources}

The last step of our study is to apply our method to other sources listed in Section \ref{sect:observations}. We have chosen several observations of five objects, GRS 1915+105, GRO J1655-40, XTE J1650-500,  XTE J1550-564 and GX 339-4, which are summarized in Table~\ref{Table2}.

At first, we examined observations of GRS 1915+105, which is a well studied source and several papers about the properties of the lightcurves with respect to the non-linear nature of the source have been published. 
We picked the observations studied by \cite{misra2004}, who classified them according to their correlation dimension into three different groups, namely those which show the chaotic behaviour (spectral classes $\beta$, $\lambda$, $\kappa$, and $\mu$), non-stochastic behaviour (spectral classes $\theta$, $\rho$, $\nu$, $\alpha$, and $\delta$), and stochastic behaviour (spectral classes $\Phi$, $\gamma$ and $\chi$). An exemplary lightcurve \oldbtxt{ taken in the class $\rho$} is shown in Figure \ref{fig:602_lc_GRS}.

In agreement with their results, we found $\bar{\mathcal{S}}>1.5$ for all observations belonging into the first and second group and we have not obtained \oldbtxt{ diagonal} lines long enough for the $K_2$ estimation for observation from class $\Phi$. 
The averaged significance ranges between 1.6 and 5.1 for different observations. 
\oldbtxt{The variability of} GRS 1915+105 is slower than in case of IGR J17091-3624, \oldbtxt{so that} the time delay $\Delta t = k {\rm d}t$\oldbtxt{is longer and ranges} between 10s to even 125s. 
Therefore, the ratio of the number of points \oldbtxt{of} the data set and the embedding delay $N/k$ is lower than for IGR J17091-3624 for similar length of observation (lower number of ``cycles'' is seen), hence lower significance could be expected for the same type of behaviour. 
The relatively high values of averaged significance thus provides quite strong evidence for the non-linear behaviour of the source.

\begin{figure}
\includegraphics[width=\columnwidth]{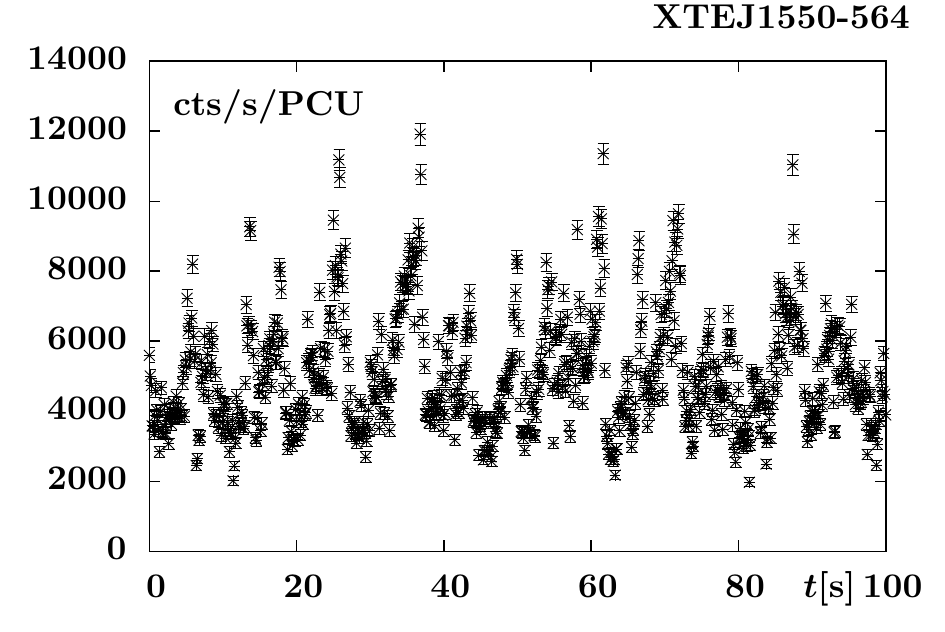}
\caption{Lightcurve of the source XTE J1550-564. The observation ID is 30188-06-03-00 and the data were extracted in the full energy range from standard-1 mode data. 
We computed the significance of the non-linear dynamics detected in this observation $\bar{\mathcal{S}}=2.89$.}
\label{fig:XTE15_lc}
\end{figure}

For the black hole candidate GX 339-4 we analysed four observations from four different spectral classes \oldbtxt{ classified} according to \cite{2012A&A...542A..56N}.
In Figure \ref{fig:GX339_lc} we show the exemplary 
lightcurve, as taken on Sep. 24, 2010, when the source was in its soft-intermediate state. Here we have found the averaged significance equal to 2.90.  For the observation taken on Mar. 26, 2010, the averaged significance is equal to 2.19.  The third observation yields $\bar{\mathcal{S}}=1.59$, which is close to our chosen threshold for the non-linear behaviour. 
We have also computed the significance with respect to the shuffled surrogates $\bar{\mathcal{S}}_{\rm shf}$, which show non-stochastic behaviour in all three cases.
For the last observation of GX 339-4, non-significant result is obtained with respect to both types of surrogates, hence the variability of the source in this state is of stochastic origin.

Figure \ref{fig:XTE15_lc} shows the lightcurve of the source XTE J1550-564, as taken on Sep 08, 1998. 
The source has been previously studied by \cite{2000ApJ...531..537S}, and the lightcurve of observation 30191-01-14-00 shown therein looks similar to the variable state of IGR J17091-3624, however on much smaller time scales. 
Thus, we needed to extract this lightcurve with very small time bin (${\rm d}t=0.032$s), which increases the influence of noise. On the other hand, this data set has much higher number of points.
Our analysis of this observation (with the highest $N/k$ ratio) yields the averaged significance $\bar{\mathcal{S}}=6.14$, which is the highest\oldbtxt{among the sample.} For comparison, we also provide the significance with respect to the shuffled surrogates, which is $\bar{\mathcal{S}}_{\rm shf}=25.17$.
Also three other observations yield quite high significance around 2.7. 
For the lightcurve presented in Figure \ref{fig:XTE15_lc}, the computed significance of the non-linear dynamics is $\bar{\mathcal{S}}=2.89$.

The other two observations of XTE J1550-564 show non-significant results ($\bar{\mathcal{S}} = 1.00$ and $\bar{\mathcal{S}}=0.30$). However, their RPs contain quite long lines compared to their shuffled surrogates, which do not show any lines longer than $\Delta t$ with the same recurrence parameters ($N_{\mathcal{S}_K}=0$ for all $\epsilon$), so that $\bar{\mathcal{S}}_{\rm shf}=3.00$ in both cases. This indicates the non-stochastic behaviour of the source during these states.

\begin{table*}[p]
\begin{tabular}{|l|c|c|c|c|c|c|c|c|c|c|c|c|}
\hline
 & & &data &state/ & &&&&{\small 20\%}&&&\\
source&ObsID&date& mode&class&N&${\rm d}t$ [s]&$m$ & $k$ &$L_{\rm max}$&$\bar{\mathcal{S}}$&$\bar{\mathcal{S}}_{\rm shf}$&$N_\epsilon$ \\
\hline \hline
GRS  &10408-01-08-00&21.05.1996&Std1&$\mu$ &289&5&8&12&438& 3.16&--&15 \\ \hline
GRS  &10408-01-10-00B&26.05.1996&SB,ch0-35&$\beta$ &6585&0.5&6&100&1439& 5.00&--&33 \\ \hline
GRS  &10408-01-10-00C&26.05.1996&SB,ch0-35&$\beta$ &6547&0.5&6&100&1685& 2.69&--&33\\ \hline
GRS  &10408-01-12-00B&05.06.1996&Std1&$\Phi$ &5466&0.5&8&50& 33& -- &--& -- \\ \hline
GRS  &10408-01-12-00C&05.06.1996&Std1&$\Phi$ &6485&0.5&8&50& 42& -- &--& -- \\ \hline
GRS  &10408-01-12-00D&05.06.1996&Std1&$\Phi$ &6626&0.5&8&50& 33& -- &--& -- \\ \hline
GRS  &10408-01-17-00&22.06.1996&Std1&$\delta$ &6812&0.5&8&60& 300& 3.17&--&17 \\ \hline
GRS&20402-01-03-00B&19.11.1996&Std1&$\rho$ &6543&0.5&8&18&458&1.63&--&33 \\ \hline
GRS&20402-01-03-00C&19.11.1996&Std1&$\rho$ &3720&0.5&8&18&372&2.47&--&34 \\ \hline
GRS&20402-01-03-00E&19.11.1996&Std1&$\rho$ &3407&0.5&8&18&309&2.68&--&30 \\ \hline
GRS  &20402-01-33-00A&18.06.1997&SB,ch0-35&$\kappa$ &6238&0.5&6&50&480& 1.63&--&27\\ \hline
GRS  &20402-01-33-00B&18.06.1997&SB,ch0-35&$\kappa$ &6690&0.5&6&50&554& 1.60&--&27\\ \hline
GRS  &20402-01-37-01A&12.07.1997&SB,ch0-35&$\lambda$ &6619&0.5&6&250&1474& 2.56&--&14\\ \hline
GRS  &20402-01-37-01B&12.07.1997&SB,ch0-35&$\lambda$ &6615&0.5&6&250&900& 2.80&--&15\\ \hline
GRS  &20402-01-45-02B&05.09.1997&SB,ch0-35&$\theta$ &6612&0.5&6&100&985& 3.29&--&27\\ \hline
GRS  &20402-01-45-02C&05.09.1997&SB,ch0-35&$\theta$ &6612&0.5&6&100&916& 5.09&--&30\\ \hline
GRS  &20402-01-45-02D&05.09.1997&SB,ch0-35&$\theta$ &6466&0.5&6&100&795& 4.10&--&23\\ \hline
GX &95409-01-12-00&26.03.2010&GE,ch5-25&HS &3170&0.5&8&5&89& 2.19&5.38&22\\ \hline
GX &95409-01-14-06&13.04.2010&GE,ch5-25&HIMS &3155&0.5&8&5& 26&0.52&0.75&8\\ \hline
GX &95409-01-16-05&29.04.2010&GE,ch5-25&SIMS &1625&1.5&8&12&66& 2.90&3.00&14\\ \hline
GX &95409-01-35-02&19.09.2010&GE,ch5-25&SS &2906&0.5&8&5&42& 1.59&2.05&8\\ \hline
XTE 1&30188-06-02-00A&07.09.1998&Std1&\oldbtxt{ HS*}&3589&0.125&7&15&92&1.00&3.00&8\\ \hline
XTE 1&30188-06-02-00B&07.09.1998&Std1&\oldbtxt{ HS*}&5019&0.125&7&15&107&0.30&3.00&15\\ \hline
XTE 1&30188-06-03-00A&08.09.1998&Std1&\oldbtxt{ HS*}&7102&0.25&8&7&115&2.89&--&25\\ \hline
XTE 1&30188-06-03-00C&08.09.1998&Std1&\oldbtxt{ HS*}&13595&0.25&8&7&125&2.76&--&25\\ \hline
XTE 1&30188-06-01-01&09.09.1998&Std1&\oldbtxt{ HS*}&10718&0.125&8&4&114&2.52&--&22\\ \hline
XTE 1&30191-01-14-00&28.09.1998&SB,ch0-17&\oldbtxt{ VHS}&109388&0.032&9&3&151&6.14&25.17&13\\ \hline
XTE 2&60113-01-13-01&19.09.2001 & Std1 &\oldbtxt{ HS}&5463&0.125&8&4&38&0.14&0.29&15\\ \hline
XTE 2&60113-01-13-02&19.09.2001& Std1 &\oldbtxt{ HS}&7127&0.125&8&4&42&0.17&0.33 &18\\ \hline
XTE 2&60113-01-18-00&24.09.2001&GE,ch5-24&\oldbtxt{ H/ST}&17715&0.1&8&4&19&-0.30&0.22&10\\ \hline
XTE 2&60113-01-18-01A&24.09.2001&GE,ch5-24&\oldbtxt{ H/ST}&19803&0.1&8&4&52&0.72&1.20&10\\ \hline
XTE 2&60113-01-18-01B&24.09.2001&GE,ch5-24&\oldbtxt{ H/ST}&15591&0.1&8&4&59&0.34&0.74&9\\ \hline
XTE 2&60113-01-18-02&24.09.2001&GE,ch5-24&\oldbtxt{ H/ST}&8581&0.1&8&4&46&0.93&1.14&10\\ \hline
XTE 2&60113-01-19-00&25.09.2001&GE,ch5-24&\oldbtxt{ H/ST}&19191&0.1&8&4&44&-0.37&-0.17&10\\ \hline
XTE 2&60113-19-01&26.09.2001 & Std1 &\oldbtxt{ SS}&1713&0.25&8&4&31&0.13&0.39&12\\ \hline
XTE 2&60113-19-04&27.09.2001 & Std1 &\oldbtxt{ SS}&1346&0.25&8&4&22&-0.28&-0.25&11\\ \hline
XTE 2&60113-29-00&17.10.2001 & Std1 &\oldbtxt{ SS}&6413&0.25&8&5&41& 0.42&0.43&14\\ \hline
GRO  &10255-01-04-00A&01.08.1996&SB,ch0-35&\oldbtxt{ VHS}&3287&0.5&7&10&1602&2.21&--&24\\ \hline
GRO  &10255-01-04-00B&01.08.1996&SB,ch0-35&\oldbtxt{ VHS}&6125&0.5&7&10&962&4.97&--&24\\ \hline
GRO  &10255-01-04-00C&01.08.1996&SB,ch0-35&\oldbtxt{ VHS}&6837&0.5&7&10&201&-0.45&3.00&18\\ \hline
GRO  &10255-01-18-00A&02.11.1996&SB,ch0-35&\oldbtxt{ VHS}&642&5&7&5&58&1.85&--&14\\ \hline
GRO  &10255-01-18-00B&02.11.1996&SB,ch0-35&\oldbtxt{ VHS}&5663&0.5&7&9&65&2.05&--&11\\ \hline
GRO  &20402-02-14-00A&28.05.1997&SB,ch0-35&\oldbtxt{ HSS}&699&5&7&10&52&1.35&3.00&12\\ \hline
GRO  &20402-02-14-00B&28.05.1997&SB,ch0-35&\oldbtxt{ HSS}&620&5&7&10&49&-0.10&3.00&11\\ \hline
GRO  &90058-16-02-00&21.02.2005&Std1&\oldbtxt{ HS*}&6502&0.5&7&10&17&0.18&1.98&3\\ \hline
GRO  &90058-16-03-00A&23.02.2005&GE,ch5-24&\oldbtxt{ HS*}&4534&0.5&7&5&32&0.22&1.74&24\\ \hline
GRO  &90058-16-03-00B&23.02.2005&GE,ch5-24&\oldbtxt{ HS*}&1973&0.5&7&5&27&0.46&1.57&17\\ \hline
GRO  &90058-16-07-00A&26.02.2005&GE,ch5-24&\oldbtxt{ HS*}&4111&0.5&7&5&36&0.17&-0.18 &23\\ \hline
GRO  &90058-16-07-00B&26.02.2005&GE,ch5-24&\oldbtxt{ HS*}&2388&0.5&7&5&38&1.86&--&21\\ \hline
GRO  &90019-02-01-00&13.03.2005&Std1&\oldbtxt{ SIMS*}&1549&0.5&6&12&45&2.33&--&5\\ \hline
\end{tabular}
\caption{Table of the RXTE observations of other sources, the meaning of the shortcuts is the following: GRS = GRS 1915+105, GX = GX 339-4, XTE 1 = XTE J1550-564, XTE 2 = XTE J1650-500, GRO = GRO J1655-40, Std1 = Standard 1 data mode, SB,ch0-35 = Single bit data mode containing energy channels 0 - 35 and GE,ch5-25 = generic event data mode using energy channels 0-25 from PCU2. The classification of the source GX 339-4 is given according to \cite{2012A&A...542A..56N} (HS = hard state, HIMS = hard-intermediate state, SIMS = soft-intermediate state, SS = soft state), the classification of the source GRS 1915+105 is given according to \cite{misra2004}, \oldbtxt{ the classification of GRO J1655-40 is according to \cite{1999ApJ...520..776S} (VHS = very high state with flaring, HSS = high/soft state) for the 1996/1997 measurement. During the 2005 outburst we used the overall plot of the hardness ratio given in \cite{0004-637X-680-2-1359} for an estimation of the spectral state (HS* - hard state, SIMS* - soft-intermediate state). Similarly we used the classification, plots and discussion given by \cite{1999ApJ...517L.121S} for XTE J1550-564 and by \cite{2003ApJ...586.1262H} for XTE J1650-500 (H/ST - hard to soft state transition).} In the 6-th and 7-th columns the used number of points of the lightcurve $N$ with the time bin ${\rm d}t$ are summarized. In the next columns, the parameters ($m,k$) od the recurrence analysis and its results -- the length of the longest diagonal line for RR$\sim$20\%, the averaged significance $\bar{\mathcal{S}}$ and the averaged significance with respect to the shuffled surrogates $\bar{\mathcal{S}}_{\rm shf}$ computed for $N_{\epsilon}$ different values of $\epsilon$ --  are presented. \label{Table2}} 
\end{table*}

\begin{figure}
\includegraphics[width=\columnwidth]{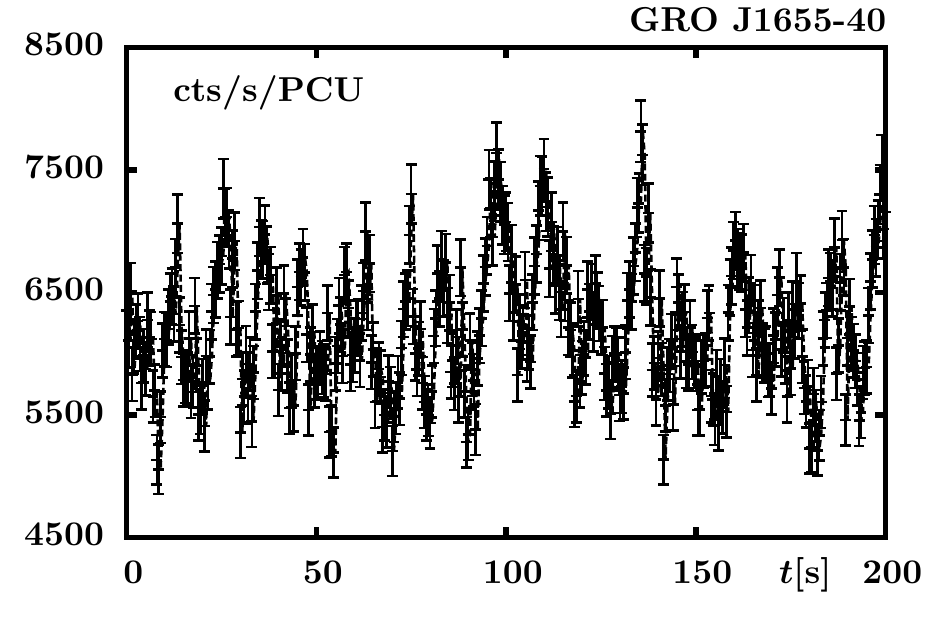}
\caption{Lightcurve of the source GRO J1655-40. The observation ID is 10255-01-04-00 and the data were extracted in the energy range 2-13 keV (channels 5-35). 
We computed the significance of the non-linear dynamics detected in this observation $\bar{\mathcal{S}}=4.97$.}
\label{fig:GRO_lc}
\end{figure}

Figure \ref{fig:GRO_lc} shows an exemplary lightcurve of the source GRO J1655-40, as taken by RXTE on Aug 01, 1996 (10255-01-04-00B). For this data set, the significance of the non-linear dynamical process that governs its variability, is above $\bar{\mathcal{S}}=4.9$.
However, in this case the high value of significance could partially be caused by the fact, that during this observation several episodes with different behaviour occur, so that the evolution is not stationary. In case of non-stationary data, the theoretical problem is much more complicated and using the surrogates created with the same spectrum may not be appropriate \citep{Theiler199277}.

Altogether, for this source we have found six observations with $\bar{\mathcal{S}}>1.5$ and seven other with non-significant results. 
For the latter group, we computed also the shuffled significance. The obsID 90058-16-07-00A shows again non-significant result, hence it is of stochastic origin. 
Three other observations (90058-16-03-00A, 90058-16-03-00B and 90058-16-02-00) yield moderate results $\bar{\mathcal{S}}_{\rm shf} \in (1.5,2.0)$, indicating \oldbtxt{a weakly} non-stochastic origin. 
The shuffled surrogates of the last three observations (10255-01-04-00C, 20402-02-14-00A and 20402-02-14-00B) have no long lines, leading to $\bar{\mathcal{S}}_{\rm shf} = 3.0$, which is a strong evidence of a non-stochastic behaviour of the source.

\begin{figure}
\includegraphics[width=\columnwidth]{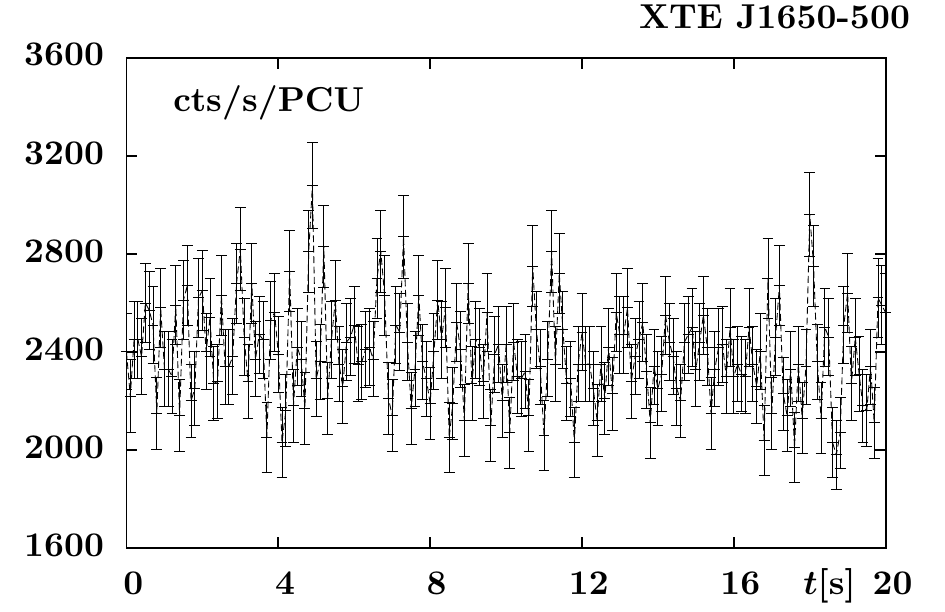}
\caption{Lightcurve of the source XTE J1650-500. The observation ID is 60113-01-18-02 and the data were extracted in the energy range 2-10 keV (channels 5-24). 
We computed the significance of the non-linear dynamics detected in this observation $\bar{\mathcal{S}}=0.93$.}
\label{fig:XTE16_lc}
\end{figure}

Finally, within the studied sample of observations of XTE J1650-500, no significant traces of non-linear behaviour have been found.
Figure \ref{fig:XTE16_lc} shows one of its exemplary lightcurves, from the observation ID 60113-01-08-02, taken on Sep 24, 2001, for which the computed significance of chaotic process was the highest. In any case, it was always below $\bar{\mathcal{S}}=1.0$. For this source also the significance with respect to the shuffled surrogates is always below  $\bar{\mathcal{S}}_{\rm shf} = 1.5$. Therefore we have not found any observation of this source, which departures from stochastic behaviour.

\section{Discussion and conclusions}
\label{sect:diss}
\subsection{Discussion} \label{subsect:diss}
In this article, we used the recurrence analysis method to study the 
non-linear behaviour of several X-ray sources. They are the black hole 
candidates, in which their X-ray luminosity originates in an accretion disk.
The latter may be responsible for a short-timescale, quasi-regular oscillations
of the emitted energy flux, through some physical process related to 
its hydrodynamical evolution.
\oldbtxt{ The possible trigger for non-linear behaviour could be the moment, when the disc
enters the parameter region, where the thermal-viscous instabilities occur.
These instabilities lead to global limit cycle oscillations of the disc configuration 
(including density, mass accretion rate, temperature, etc.).
Such unstable evolution of the disc translates into non-linear patterns in the light curve. }

Two microquasars, GRS 1915+105, and recently discovered IGR J17091-3624,
are already firm candidates for the limit-cycle oscillations that are called also the 'heartbeat' state (see e.g., \citealt{2011ApJ...737...69N}). Our present analysis confirmed that the variability in these sources is significantly governed by the non-linear dynamics of accretion process.
Other sources of that kind, which accrete at presumably high accretion rates
and exhibit soft, disk-dominated, states, have not yet been studied 
extensively enough to tackle this problem quantitatively.
The question however remained, whether GRS 1915+105, and later its analogue, IGR J17091-3624, are the only two sources of this kind, or rather such behaviour 
can be common to all accreting black holes, if only the physical conditions
which trigger the non-linearity, are similar in there.

\oldbtxt{ For other X-ray binaries than these two microquasars, the situation is not so obvious. However, \citet{2011MNRAS.414.2186J} discussed the theoretical aspects of the thermal-viscous instabilities in the accretion disks
and presented observational constraints for a sample of Galactic black hole systems. In particular, for the radiation pressure instability, a few other sources were hinted, based on their estimated Eddington ratios and observed properties. It should be stressed, that the accretion rate at exactly Eddington rate is not required by the theory for the radiation pressure instability to develop. 
Therefore we aim to search for the non-linear hydrodynamics of the accretion disk hidden in the data also for these other sources.
}

The black hole X-ray transient source XTE J1650-500 has shown its only 
outburst in the year 2001, and there were not many observations available 
for that source.
\citet{2003ApJ...586.1262H} detected the high-frequency variability in
XTE J1650-500 and interpreted them as the orbital 
frequency at the innermost stable orbit around a Schwarzschild black hole, while
\citet{2003ApJ...586..419K} suggested that the accretion flow geometry is that of the disk-jet-corona system.
There is not firm estimation in the literature about the value of the accretion rate in this 
source.
\citet{2003ApJ...592.1100T} detected in this source the X-ray flares which 
have  durations between 62 and 215 s and peak fluxes that are 5-24 times 
higher than the persistent flux. Therefore \citet{2011MNRAS.414.2186J} tentatively 
hinted at this X-ray binary as a prospective candidate for the disk instability 
induced oscillations, because of these timescales. 
These flares have however non-thermal energy 
spectra, and possibly do not originate in the accretion disk/corona
 at all. 
 \oldbtxt{ They also occur at luminosities about thousand times lower than 
 the peak outburst luminosity, probably at the level about $2 \cdot 10^{-5}L_{\rm Edd}$.
 Therefore they are not good candidates for the radiation pressure instability. 
 \citet{2003ApJ...592.1100T} also reported about aperiodic oscillation with 14 days time scale.
 They related these variations of X-ray light curve with the ionization instability. However such a
 long variability could not be studied by our present method, because no continuous observation
 of required length (at least one order of magnitude longer than the period) is available. }
Our current analysis was based on the earlier observations, from up to October 2001, with a higher count rate. It 
showed that in this source the significance of the 
underlying non-linear dynamics is very low.
We would therefore rather reject it as a candidate for the limit-cycle type of 
oscillations. 
This source requires further monitoring to put more firm conclusions about 
the nature of its variability.

For the other sources in our study, the significance of the underlying 
deterministic chaotic process, being the intrinsic to the accretion disk 
reason for its observed variability, is much higher. This result is in full 
agreement with the observational facts, noted during the time of over 20 
years of collecting data by the X-ray satellites.

In XTE J1550-564, at the peak of the outburst the luminosity is close to $L_{\rm Edd}$ \citep{2011MNRAS.411L..66H}, and the quasi-periodic oscillation (QPO) observed in this source during its 1998 outburst were found to be linked to the disc count rate \citep{2000ApJ...531..537S}. XTE J1550-564 is classified as a microquasar on the basis of its large-scaled moving jets, detected at X-ray and radio \citep{2002Sci...298..196C}, similarly with GX 339-4 \citep{2002ApJ...573L..35C}. We expect that these two objects should have similar characteristics of the disk variability, and along with the two well studied microquasars the non-linear dynamical processes in these sources should also occur. Our current analysis does confirm these expectations.

The spectral state of GRO J1655-40 was initially 
not determined (i.e., the low, high, or very high state) for its first outburst in 1994, based on the BATSE observation 
\citep{1996ApJ...463L..79C}. However, \citet{1999ApJ...522..397R} after the analysis of the data spanning the X-ray outburst from 1996-1997, estimated the unabsorbed X-ray luminosity to above $\sim 0.2 L_{\rm Edd}$ during the peak. Such luminosity and accretion rate
 is quite sufficient for the 
instability of the accretion disk, which develops within the innermost 60-80 
Schwarzschild radii from the black hole, and leads to moderate luminosity oscillations \citep{2011MNRAS.414.2186J}. 
\citet{1998ApJ...499L.187M} studied the evolution of GRO J1655-40
through the high, intermediate, and low state. This source at the beginning of its decay might have even shown signatures of a very high state, just like other black hole candidates.
\citet{1999ApJ...520..776S} presented the full spectral analysis of the RXTE data for 1996-1997 outburst of GRO J1655-40 and showed that during the high/soft state, its spectrum is dominated by the soft thermal emission from the accretion disk. 
Comparing the two above mentioned black hole binaries, 
\citet{2000ApJ...531..537S} found that in both sources the QPO frequency is 
correlated with the disk flux, and hence with the rate of mass accretion 
through the disk, confirming therefore that the intrinsic mechanism of this variability is related to the accretion process. Most recently, 
\citet{2015arXiv150408313U} reported about an unusual soft state of GRO J1655-40, observed during its 2005 outburst by the RXTE. Chandra X-ray grating observations have revealed a high mass-outflow accretion disc wind in this state, which can be related to the accretion disk (in)stability evolution
\citep{2015A&A...574A..92J}. Indeed, \citet{0004-637X-680-2-1359} showed, that during its 2005 outburst, GRO J1655-40 ejected massive winds, possibly driven by magnetic process in the accretion disk.

\oldbtxt{ The fact, that we have found non-linear dynamics hidden in the light curves of almost all the studied sources means, 
that the evolution of the accretion disc, and possibly corona, is important for the outgoing radiation and that the accretion flows is influenced by a physical instability, at least at some specific states.
}

\oldbtxt{In App.~\ref{sect:poincare} we summarized the results presented in \citep{chaos_proc}, where we test the capabilities of our method on two numerical trajectories. We showed how the method works with simulated trajectories of a complicated non-linear system (motion of the test particle in the field of a black hole surrounded by thin massive disc given by exact solution of Einstein equations), from which one is a regular orbit, while the other one is chaotic. The chaotic orbit shows higher significance of non-linear dynamics. We studied the influence of noise on the output of the method. While the significance for regular orbit quickly drops below unity, the significance of chaotic motion remains high up to comparable variance of the noise and the data.}

\subsection{Conclusions}
\begin{itemize}
\item We applied the recurrence analysis to the observations of six black hole  X-ray binaries detected by RXTE satellite.
\item We developed a method for distinguishing between stochastic, non-stochastic linear and non-linear processes using the comparison of the quantification of the recurrence matrices for the real and surrogate data.
\item We tested our method on the sample of observations of the microquasar IGR J17091-3624, which spectral states were provided by \cite{0004-637X-783-2-141}. Results \oldbtxt{with $\bar{\mathcal{S}}>1.5$} for the ``heartbeat'' variable $\rho$ state were obtained, \oldbtxt{which indicates the non-linear dynamics. Hence, our results also corroborate the validity of simulations of radiation pressure instability induced oscillation of the accretion disc in this source reported by \cite{2015A&A...574A..92J}.}
\item We examined several observations of other five microquasars. Aside from the well-studied binary GRS 1915+105, we found significant traces of non-linear dynamics also in three sources (GX 339-4, XTE J1550-564 and GRO J1655-40). 
\item \oldbtxt{ The non-linear behaviour of the lightcurve during some observations gives the evidence, that the \oldbtxt{temporal evolution of the} accretion flow in the binaries is governed by 
\oldbtxt{low-dimensional system of non-linear equations and the dynamics is chaotic.  Hence the variability in the lightcurve is driven by a dynamical system of the accreting gas described by a few parameters (e.g., the accretion rate) rather than by stochastic variations coming from random processes (flares) in the disc.} 
Possible explanation is that the accretion disc is in the state prone to the thermal-viscous instability and is undergoing the induced limit cycle oscillations.}
\item \oldbtxt{As pointed out in \cite{misra2004} the knowledge about the chaotic nature of the system can enlighten the fundamentals of the physical processes yielding the observed radiation. The rough estimate of the R\'enyi's entropy, which is related to the maximal Lyapunov exponent, can be compared with the estimates for magnetohydrodynamical simulations of the accretion flow to further confirm or refute the accretion models.}
\end{itemize}

\oldbtxt{Our analysis of course does not answer the question of how the appearance or absence of non-linear chaotic variability is related to the evolution of a given X-ray binary during its outburst and the spectral state transitions. Such analysis is extremely complex and currently beyond the scope of this article, but definitely worth further studies. For now, our tentative picture would be that 
the purely stochastic variability occurs when the source is in its hard state, at the very beginning of the outburst. 
In the disk dominated soft state, as well as in the intermediate states, the non-linear variability due to chaotic process in the accretion disk may appear. 
However, even in this state, the intrinsic coronal emission, which should be rather stochastic, as well as possibly reprocession of the hard X-ray flux in the disk, may significantly affect the observed flux and hence our results. 
On the other hand, if the corona above the disk forms through the evaporation of the upper layers of an accretion disk, then its variability will be related to the disk variability, with a possible short time delay \citep{2005MNRAS.356..205J}. 
}

\section*{Acknowledgments}
We thank Ranjeev Misra,  Piotr Zycki, Fiamma Capitanio and Bozena Czerny
 for helpful discussions. 
\oldbtxt{  We also thank to the anonymous referee, whose comments helped us to make the paper better organised and more clear for the readers.}
This work was supported in part by the grant DEC-2012/05/E/ST9/03914 from the
Polish National Science Center.


\bibliographystyle{aa} 
\bibliography{Sukova} 


\begin{appendix}

\section{Recurrence analysis of the observed time series of IGR~J17091-3624} 
\label{sect:Recurrence}

\begin{figure}
\includegraphics[width=\columnwidth]{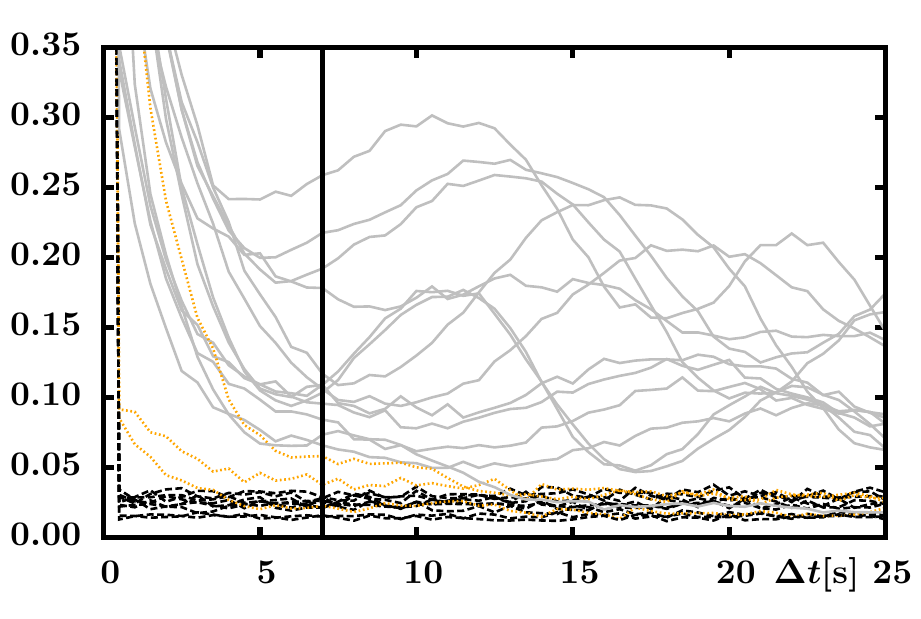}
\caption{The dependence of mutual information on the time delay $\Delta t$ for the set of observations of IGR J17091-3624 listed in Table~\ref{Table1}. \oldbtxt{Observations with low and constant mutual information are} plotted with black dashed line, \oldbtxt{observations with oscillating mutual information are} plotted with grey solid line and \oldbtxt{those with intermediate behaviour are} plotted by yellow dotted line. The vertical line indicates the chosen time delay for our analysis.
}
\label{fig:fig1}
\end{figure}

The recurrence analysis is based on the study of the times, when the trajectory returns close to itself in the phase space (closer than a certain recurrence threshold $\epsilon$). However, the trajectory has to be reconstructed from the observed time series with the time delay technique. Hence the resulting phase space vector is given as
\begin{equation}
\vec{y}(t) = \{x(t),x(t+\Delta t),x(t+2\Delta t),\dots,x(t+(m-1)\Delta t)\},
\end{equation}
where $x(t)$ is the time series, $\Delta t$ is the embedding delay and $m$ is the embedding dimension. 
When working with measured time series, the embedding delay $\Delta t$ has to be an integer multiple of the binning time of the data ${\rm d}t$, hence  $\Delta t = k \,{\rm d}t$.
The determination of these parameters is a subtle issue \citep{PhysRevA.45.3403,cao1997practical,Thiel2004_1_1667633,Marwan2007237}. 
The usual practise is to determine the time delay as the first distinct minimum of the time delayed mutual information, for which we use the procedure {\tt mutual} from the {\tt TISEAN} package. 
In Fig.~\ref{fig:fig1} 	the dependence of the mutual information on the time delay $\Delta t$ is given for the set of observations of IGR J17091-3624. 
The observations \oldbtxt{show two kinds of behaviour.}
Mutual information of some lightcurves is very low even for small time delays and is roughly constant for all values of $\Delta t$ (black dotted lines), whereas the course of mutual information of others is smooth and shows oscillations with period ranging from $10\jd{s}$ to $20\jd{s}$ (grey solid lines). 
Three observations show intermediate behaviour, there is a smooth decrease of the mutual information, but no clear second maximum, the values then rather stabilize around a constant value of the first minimum (yellow dotted line). 
Because we want to compare the results of the analysis with the same parameters for all observations, we choose the time delay to be $\Delta t = 7 \jd{s}$ (for ${\rm d}t=0.5\jd{s}$ is $k=14$) as a reasonable value for each lightcurve.

\begin{figure}[t]
\includegraphics[width=\columnwidth]{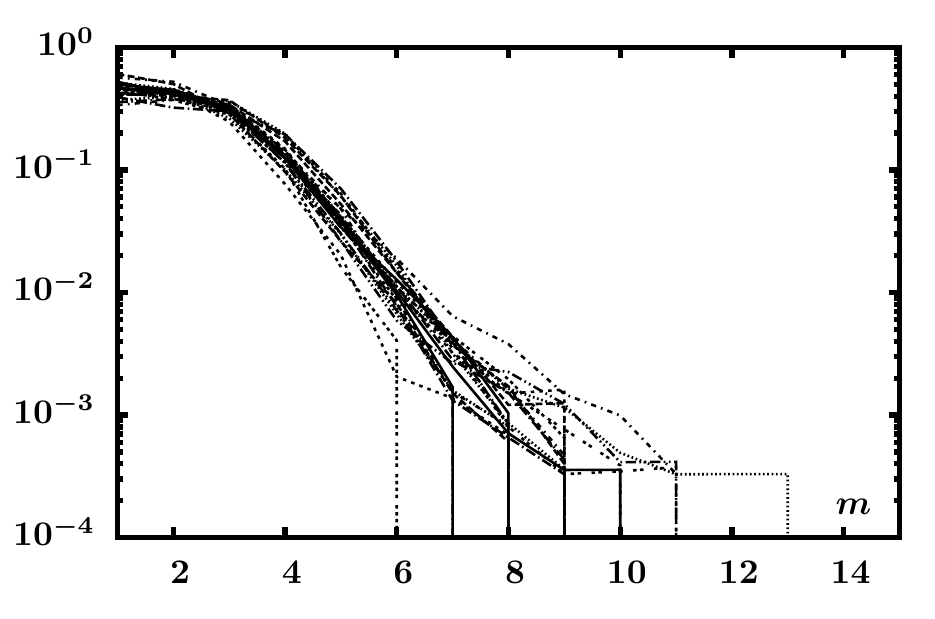}
\caption{The dependence of the ratio of false nearest neighbours on the embedding dimension $m$ for the set of observations of IGR J17091-3624 for the choice $R_{\rm tol}=10$.
}
\label{fig:fig2}
\end{figure}

\begin{figure*}
\includegraphics[width=\columnwidth,height=0.85\columnwidth]{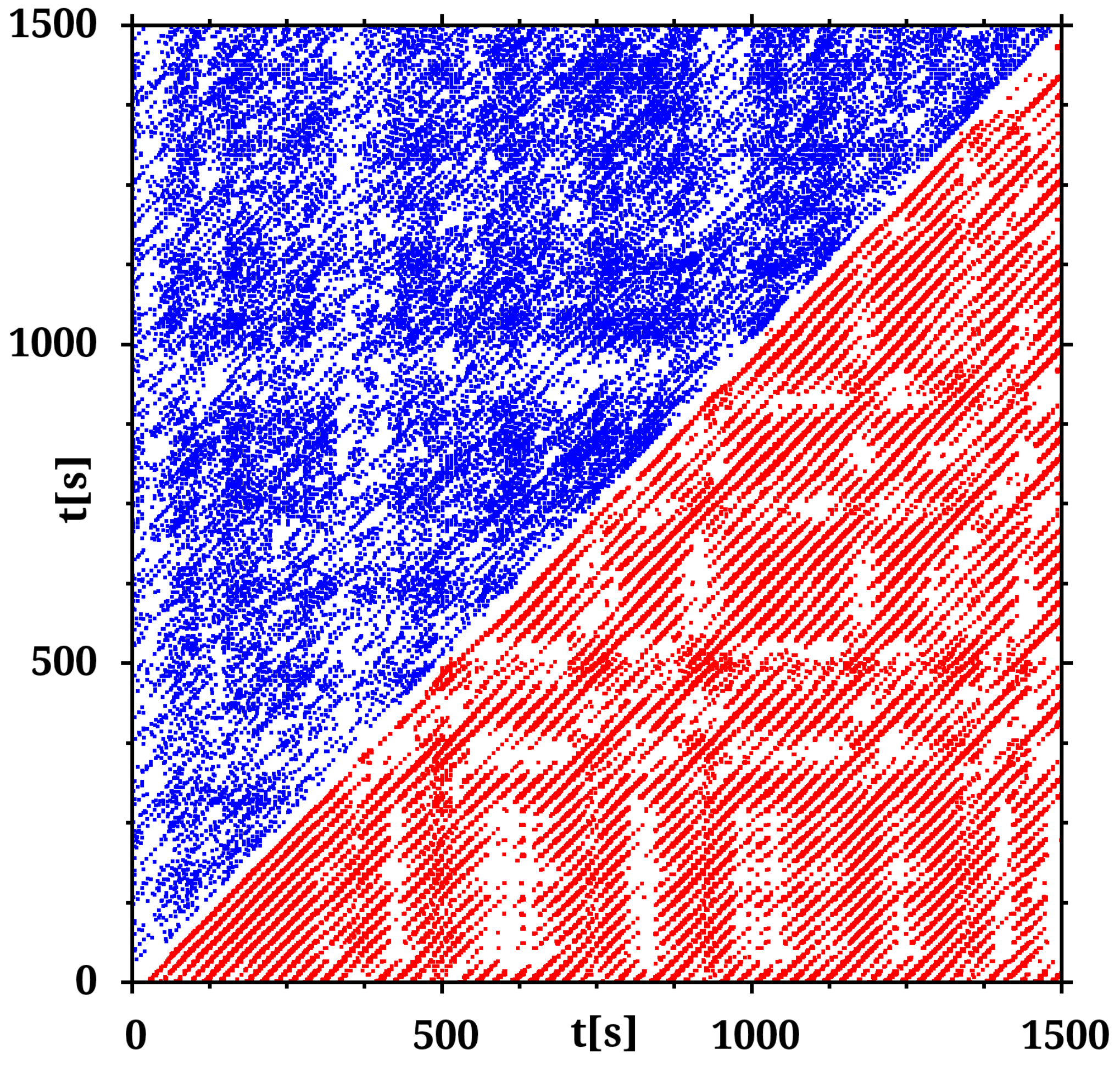}
\includegraphics[width=\columnwidth,height=0.85\columnwidth]{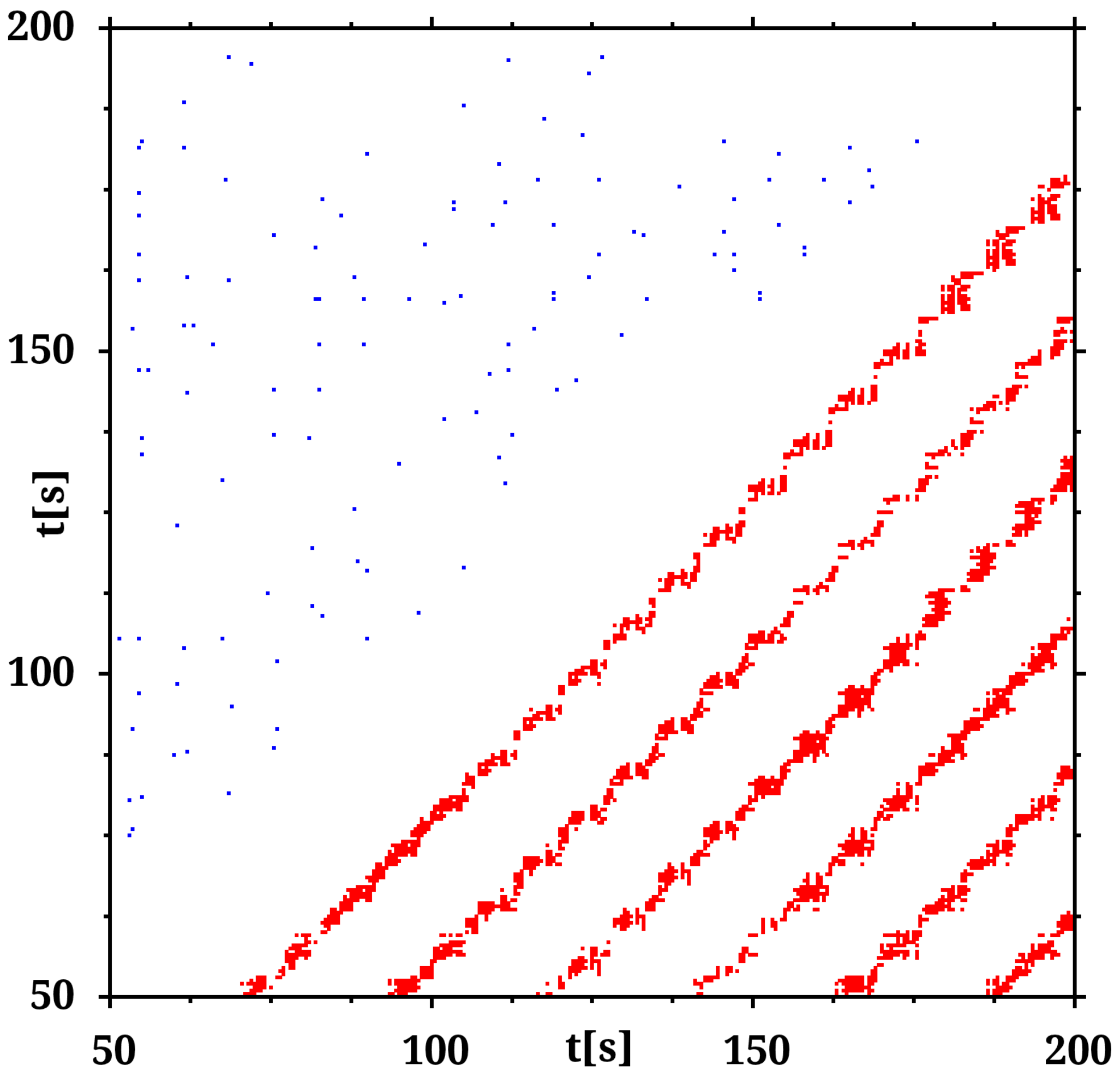}

\includegraphics[width=\columnwidth,height=0.85\columnwidth]{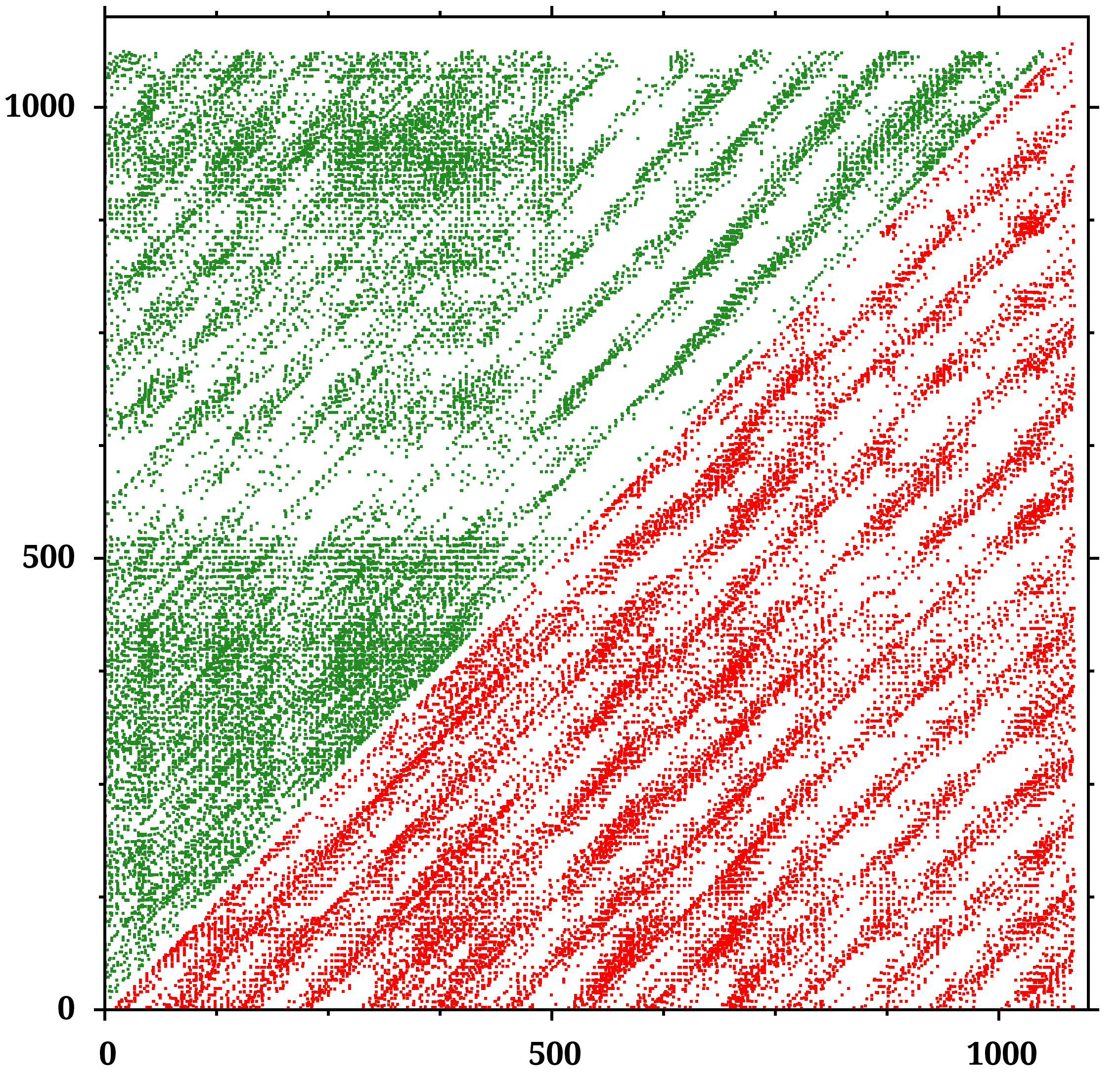}
\includegraphics[width=\columnwidth,height=0.85\columnwidth]{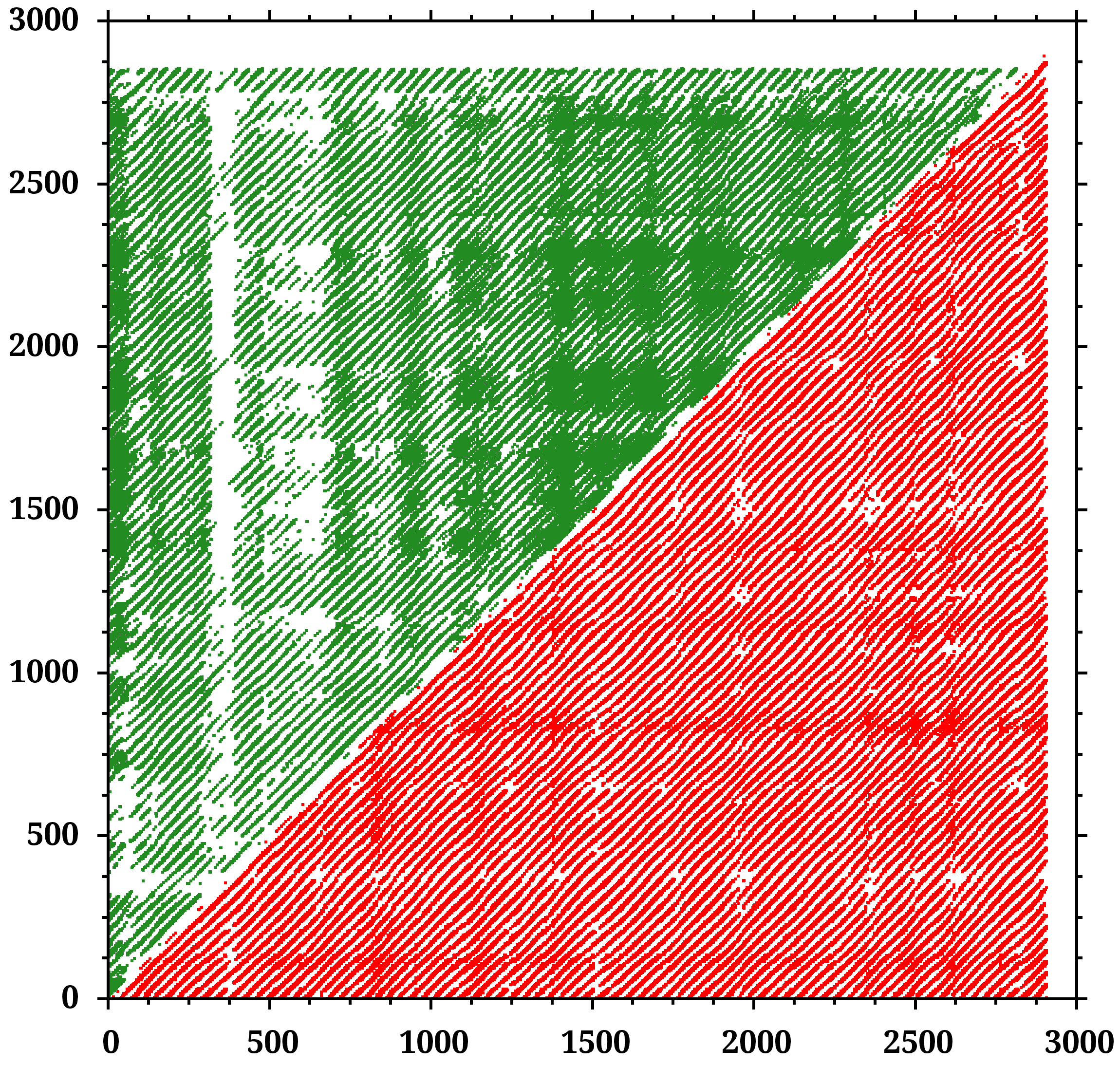}

\includegraphics[width=\columnwidth,height=0.85\columnwidth]{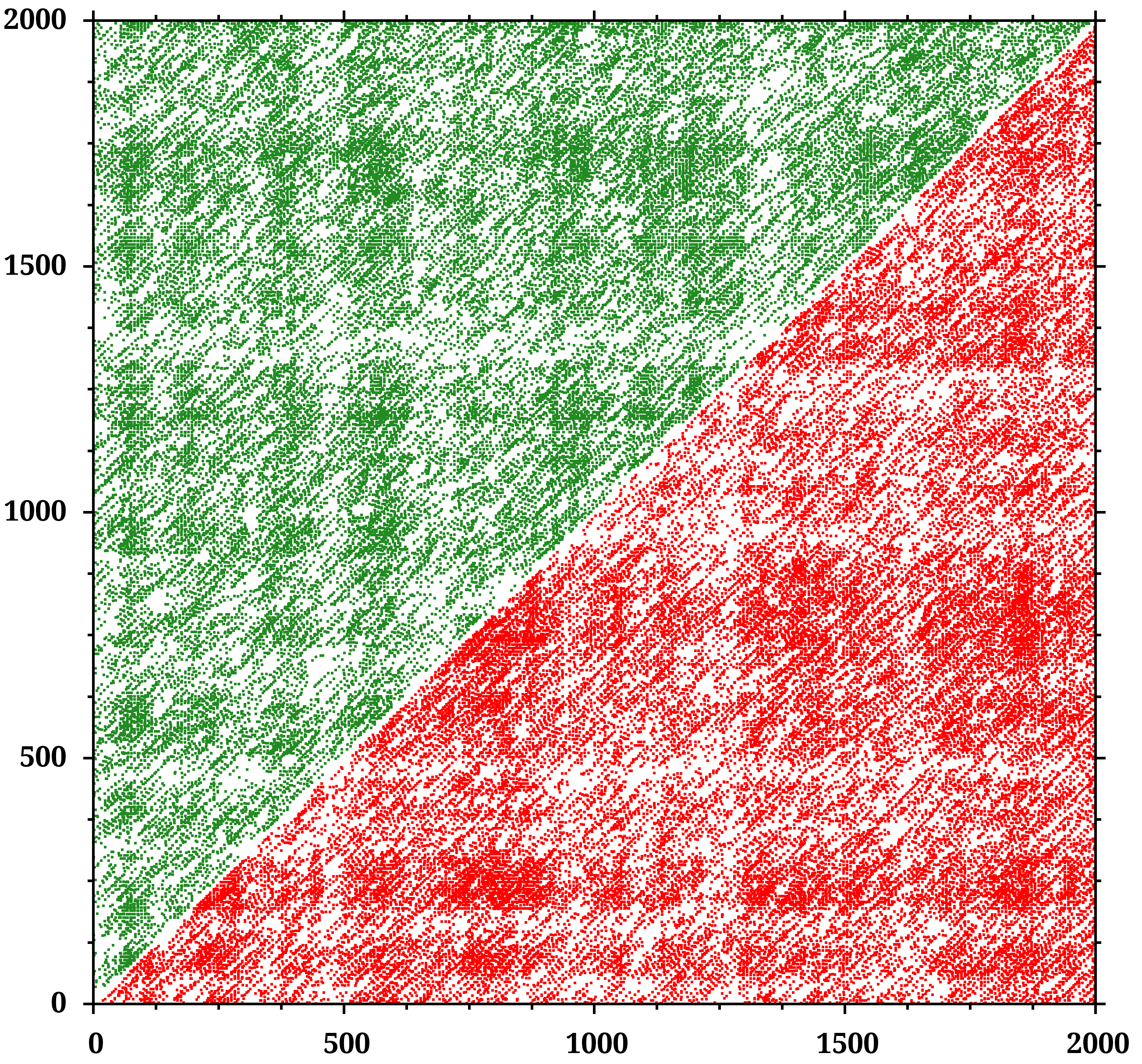}
\includegraphics[width=\columnwidth,height=0.85\columnwidth]{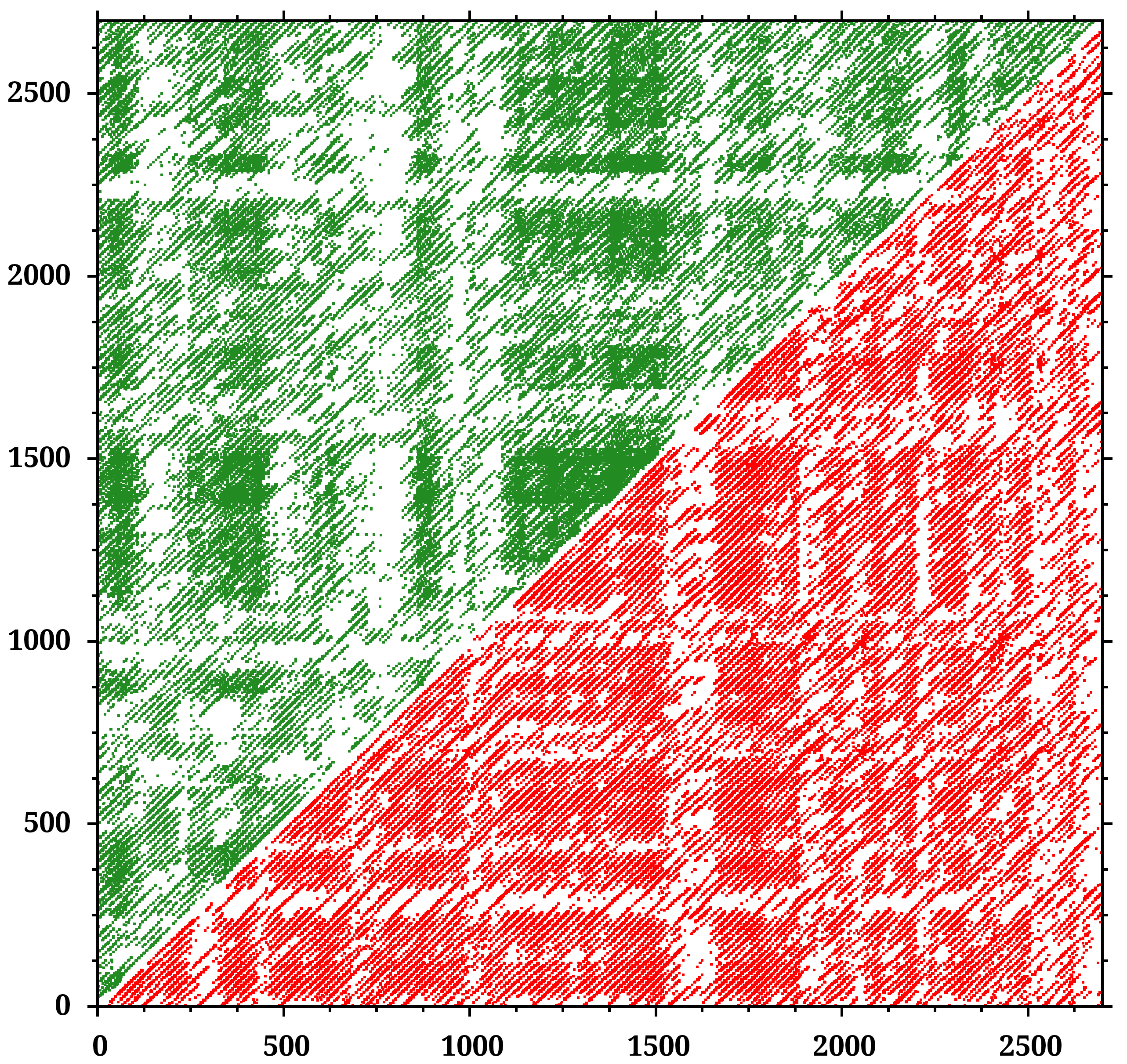}
\caption{RP for several observations \oldbtxt{of IGR J17091-3624 from Table~\ref{Table1}.} 1. row: left -  SIMS/HIMS state 02-00 (left upper corner in blue) and $\rho$ state 06-02 (right lower corner in red). Long diagonal lines appear in the red RP. Right - The zoom of preceding plot. The diagonal lines are affected by the noise in such a way, that the apparently long lines are composed of shorter square-shaped pieces.\oldbtxt{Next rows:} In bottom red halves of each plot RP is plotted for observations 04-01 (2. row left, IVS state), 04-02 (2. row right, $\rho$-state), 05-02 (3. row left, SIMS state) and 07-01 (3. row right, $\rho$-state). In green colour in the left upper half of each plot one of the corresponding surrogates is shown.
}
\label{fig:fig4}
\end{figure*}

For finding the appropriate embedding dimension, the well known method is to determine the number of false nearest neighbours \citep{PhysRevA.45.3403}. 
This procedure is based on the fact, that for too low embedding dimension the points, which are not close on the attractor, can be projected into the smaller space as very close neighbours. 
When the dimension is increased, the ambiguity of their mutual position is removed and their distance grows substantially.  
Hence, we set a tolerance threshold $R_{\tt tol}$ and denote the point as false neighbour, if the distance in $m$ and $m+1$ dimension increase by a factor bigger than $R_{\tt tol}$. 
In this way we investigate the change of distance for each point of the trajectory and its nearest neighbour in the reconstructed phase space.  
If the dimension is big enough to cover the whole attractor, the fraction of false neighbours goes to zero. 
We use the procedure {\tt false\_nearest} from the {\tt TISEAN} package. 
However, because our data are both short and noisy, we have to skip the second criterion, which omits points being farther than the standard deviation of the data divided by the threshold. 
Otherwise we would have very small number of points to consider and the results are meaningless. 
Therefore we are forced to include also points, which are not very close neighbours (even when they are the nearest ones), because due to the noise and the limited amount of data, there is not enough really close points. 
We also have to be careful about the size of the threshold $R_{\tt tol}$, which cannot be very high, otherwise almost no nearest neighbour distance can grow by this factor due to a finite size of the attractor.
This can slightly affect the results. 

We choose the threshold $R_{\tt tol} = 10$. The corresponding false nearest neighbour ratio for the set of IGR's observations is given in Fig.~\ref{fig:fig2}. 
We set the embedding dimension $m=10$, in which for almost all observations the false nearest neighbour ratio drops to zero. 
However, even $m=6$ could be a good choice, because for this value there is only 1\% or less of false neighbours.

These two methods are used for guessing some appropriate values of the two parameters, but the results of the recurrence analysis do not strongly depend on it. This is supported by \cite{Thiel2004_1_1667633}, who claimed that some of the dynamical invariants estimated by recurrence analysis including second order R\'enyi entropy $K_2$ do not depend on those parameters.

\begin{figure}[t]

\includegraphics[width=\columnwidth]{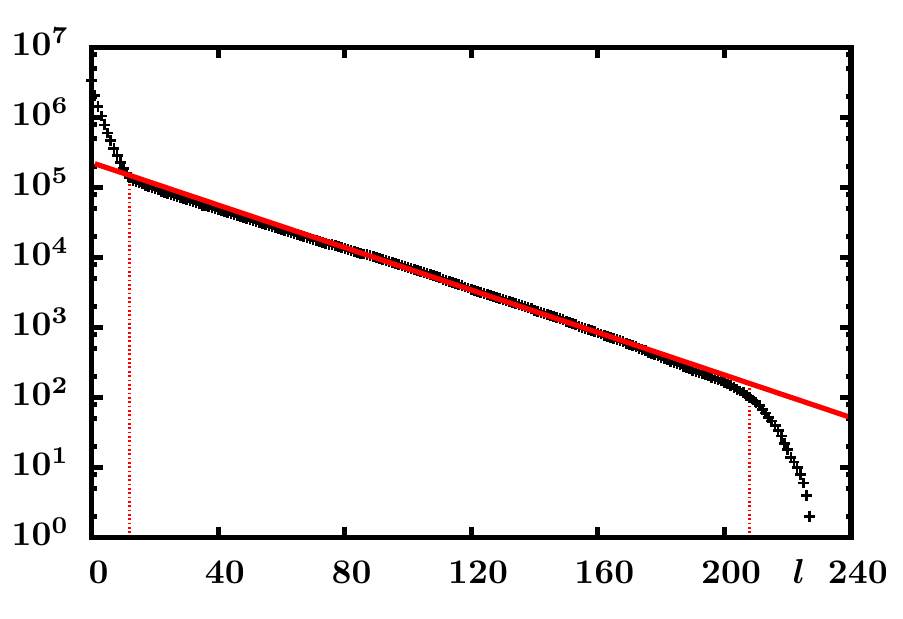}
\caption{ The cumulative histogram $p_c(\epsilon,l)$ for the observation 06-02 computed for $\epsilon=2.8$. The length of the lines $l$ is given in number of points of the lines (thus the total time of the line is given as $t_l = l \Delta t$). The estimate of the  R\' enyi's entropy $K_2=0.0699\pm0.0003$ is obtained as the linear regression of the central part of the cumulative histogram (indicated by solid red line, the dashed red lines mark the used part of the histogram).
}
\label{fig:602histogram}
\end{figure}

Having the values for embedding delay and dimension determined we can now proceed with the recurrence analysis. 
The basic object of the analysis is the recurrence matrix defined as follows:
\begin{equation}
\mathbf{R}_{i,j}(\epsilon) = \Theta (\epsilon - \parallel \vec{x}_i - \vec{x}_j \parallel ), \qquad i,j = 1,...,N, \label{RP_def}
\end{equation}
where $\vec{x}_i = \vec{x}(t_i)$ are ($N$) points of the reconstructed phase trajectory and $\Theta$ is the Heaviside step function. The matrix thus contains only 1's and 0's and can be easily visualised by plotting a dot at the coordinates $i$, $j$ whenever \oldbtxt{ R}$_{i,j}(\epsilon)=1$, which is called the recurrence plot (RP). From the definition (\ref{RP_def}) it is obvious, that the RP is symmetrical with respect to the main diagonal, which is formed by dots (for $i=j$) and which is omitted in further computations.

The comparison of the RPs for two observations, 02-00  (SIMS/HIMS -- upper left corner on the left panel in first row) and 06-02 ($\rho$-class -- lower right corner of the same panel) is given in~Fig.~\ref{fig:fig4}. 
Because the plot is symmetrical  with respect to the main diagonal, we show only one half of the RP for each observation in the upper/bottom trilateral half of the plot. 
Different geometrical structures contained in the RP could be seen in these two examples. The structures can be quantified based on the statistical properties of the recurrence matrix.

The most important entity for the recurrence analysis is the existence of long diagonal lines, which indicate periodic (and regular) behaviour. 
Such structures are much more prominent in the RPs of $\rho$ observations. 
The quantification of the visual information is contained in the histogram of diagonal lines of a certain prescribed length $l$,
\begin{equation}
P(\epsilon, l) = \sum_{i,j=1}^N(1-R_{i-1,j-1}(\epsilon))(1-R_{i+l,j+l}(\epsilon))\prod_{k=0}^{l-1}R_{i+k,j+k}(\epsilon). 
\end{equation}
Based on the histogram more other quantifiers can be defined \citep{Marwan2007237}.
Here we will focus mainly on $L_{\rm max}={\rm max}(\{l_i\}_{i=1}^N)$, which is the length of the longest diagonal line (except of the main diagonal), and on the estimate of the second-order R\' enyi's entropy $K_2$.
This entropy, also called correlation entropy or correlation exponent, collision entropy or just R\' enyi's entropy  \citep{Grassberger1983227}, is defined for the phase trajectory $\vec{x}(t)$ as
\begin{equation}
K_2 = - \lim_{\Delta t \to 0} \lim_{\epsilon \to 0} \lim_{\Delta l \to \infty}  \frac{1}{l\Delta t} \ln \sum_{i_1,\dots,i_l} p_{i_1,\dots,i_l}^2(\epsilon), \label{K2_def}
\end{equation}
where the phase space is coarse-grained into $m$-dimensional hyper-cubes of size $\epsilon$. Here,
$p_{i_1,\dots,i_l}$ describes the probability that the first point of the trajectory $\vec{x}(t=\Delta t)$ is inside the box $i_1$, the second point $\vec{x}(t=2 \Delta t)$ is in the box $i_2$ and so forth to the $l$-th point being in the box $i_l$. It can be shown  \citep{Marwan2007237}, that the square of this probability is connected with the existence of a diagonal line in the RP (the diagonal line means, that the trajectory goes twice through the same sequence of phase-space boxes).

\begin{figure}[t]

\includegraphics[width=\columnwidth]{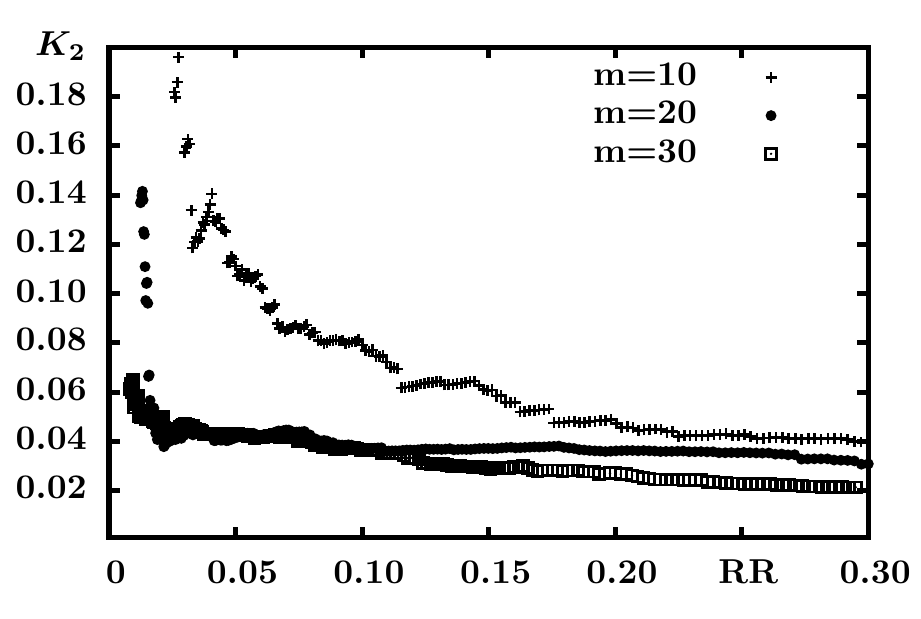}
\caption{ R\' enyi's entropy $K_2$ versus recurrence rate for the observation 06-02 for three different embedding dimensions marked in the figure.
}
\label{fig:602K2}
\end{figure}

More precisely, $K_2$ is related with the cumulative histogram of diagonal lines $p_c(\epsilon,l)$, describing the probability of finding a line of minimal length $l$ in the RP, by the relation
\begin{equation}
p_c(\epsilon,l) \sim \epsilon^{D_2} e^{-l \Delta t K_2}, \label{cumul}
\end{equation}
hence we can estimate the value of $K_2$ as the slope of the logarithm of the cumulative histogram versus $l$ for constant $\epsilon$ (see an example in~Fig.~\ref{fig:602histogram}).
This estimation holds for the limit $l\to\infty$, particularly for $l>\Delta t$.
The important property of $K_2$ entropy is that it is the lower estimate of the sum of positive Lyapunov exponents of the system, hence it is positive for chaotic systems \oldbtxt{\citep{Marwan2007237}}.

Considering formula (\ref{cumul}) for two different values of threshold $\epsilon_2=\epsilon_1 + \Delta \epsilon$, we can also estimate the value of the correlation dimension $D_2$ by
\begin{equation}
D_2(\epsilon)=\ln \left( \frac{p_c(\epsilon_1,l)}{p_c(\epsilon_2,l)} \right) \Bigg/ \left( \frac{\epsilon_1}{\epsilon_2} \right).
\end{equation}
However, for obtaining reasonable results for $D_2$ much more data points are needed than for estimating $K_2$ \citep{Marwan2007237}. 
Because we are limited by the length of individual observations, which are not longer than several thousand seconds yielding twice data points for our chosen binning time ${\rm d}t = 0.5s$, we were unable to determine the value of $D_2$ by means of the recurrence analysis.

\begin{figure}[t]
\includegraphics[width=\columnwidth]{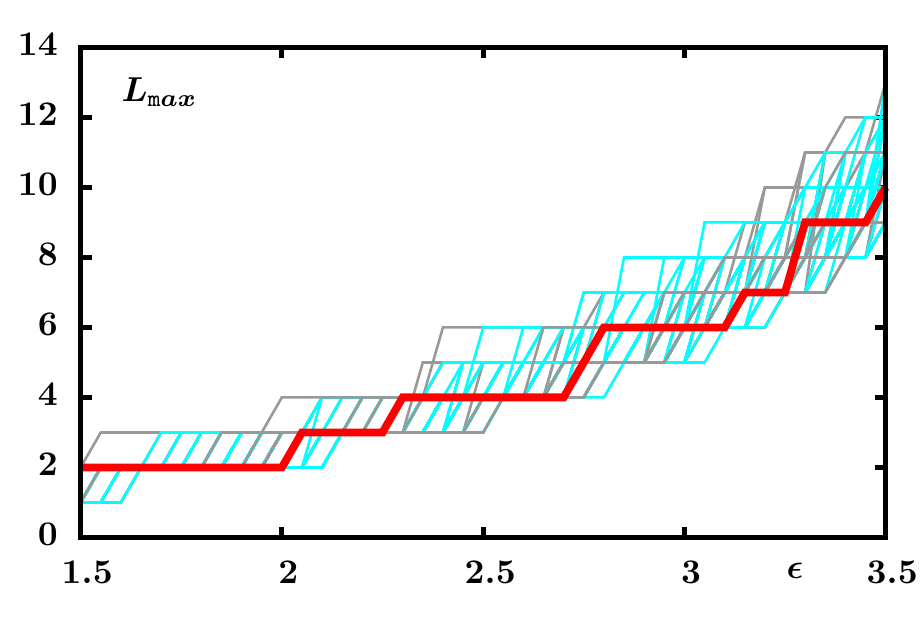}
\caption{ Dependence of the longest diagonal line $L_{\rm max}$ given in points on the recurrence threshold $\epsilon$ for the observation 05-02  (thick red line). The ensemble of 100 surrogates are shown in grey and the ensemble of shuffled surrogates are shown in cyan.
}
\label{fig:502Lmax}
\end{figure}

Besides the short \oldbtxt{duration} of the observation, another difficulty is the presence of noise, which affects the structures contained in RP. 
This is documented by the right panel in first row of Fig.~\ref{fig:fig4}, which shows the zoom of the left panel for $t\in(50,200)\jd{s}$.
Here the apparently long lines in fact consist of shorter and broader pieces separated by gaps. 
The gaps appear, when the noise takes the ``should-be recurrence point'' out from the $\epsilon$-neighbourhood, and on the other hand, the edges on the lines appears, when noise brings the ``should-not-be recurrence point'' inside.  
Hence, there is more shorter diagonal lines and less longer ones, thus the slope of the histogram is steeper.

The most usual way to deal with the noise in data is to take the radius of the neighbourhood big enough, so that it covers the noisy perturbations of the position, which according to \cite{thiel2002influence, Marwan2007237} is $\epsilon = 5\sigma$, where $\sigma$ is the standard deviation of the noise.
However, a part of the noise is caused by the measurement process and another part stems from the stochastic processes behind the emission of the radiation. 
Moreover, the amount of noise and its standard deviation is unknown, but it is quite large comparing to the standard deviation of the whole data set (see Fig.~\ref{fig:602_lc_IGR}). 
Hence, such a choice of threshold would not be appropriate, because it would be huge and almost every point would become the recurrence point.

We consider such thresholds, which yields the recurrence rate (RR -- ratio of the recurrence points to all points of the recurrence matrix) within 1\% to 25\% and we have to keep in mind, that due to the noise, the slope of the cumulative histogram and $K_2$ is overestimated. 
An example of the decreasing dependence of the $K_2$-estimate  on the recurrence rate with $\Delta t = 7\jd{s}$, and three different embedding dimensions $m_1=10$, $m_2=20$ and $m_3=30$, is given in Fig.~\ref{fig:602K2}. 

\section{Comparison between the surrogate data and observations of IGR J17091}
\label{Sect:Comparison}

\begin{figure}[tb]
\includegraphics[width=\columnwidth]{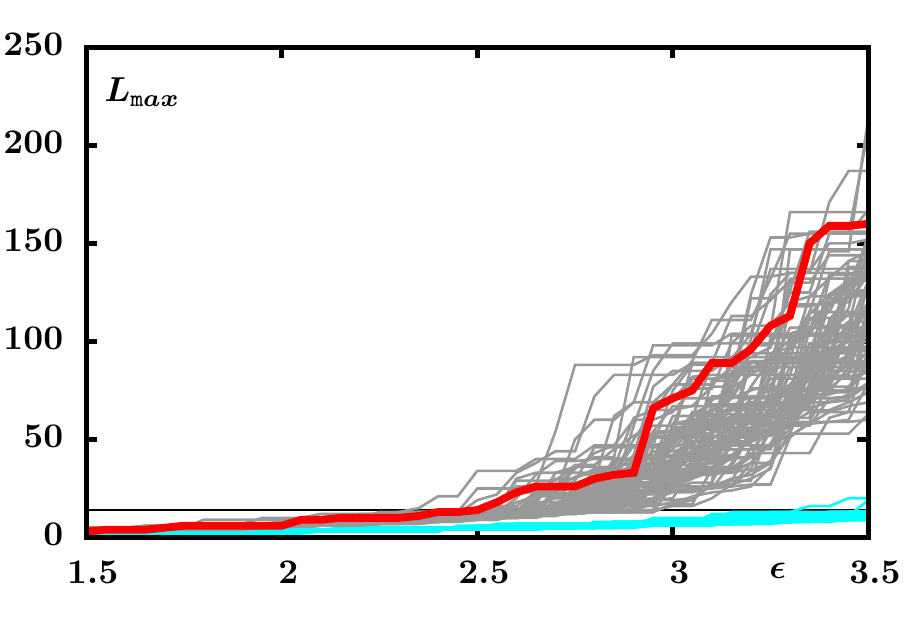}
\caption{ The same as in Fig.~\ref{fig:502Lmax} for the observation 05-01. The condition $l>\Delta t$ on lines' length used for $K_2$-estimation is indicated by black horizontal line.
}
\label{fig:501Lmax}
\end{figure}

\begin{figure}[t]
\includegraphics[width=\columnwidth]{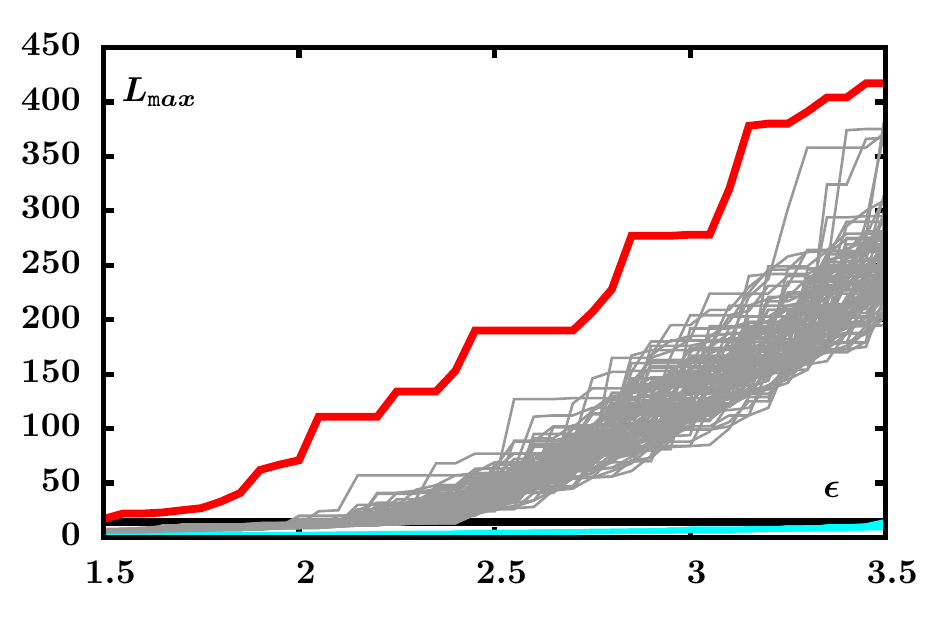}
\caption{ The same as Fig.~\ref{fig:502Lmax}, but for the observation 06-02.}
\label{fig:602Lmax}
\end{figure}



We first study the appearance of long diagonal lines in RP. For the real data set and its surrogates we compare the dependence of $L_{\rm max}$ on chosen $\epsilon$. 
We obtain three different types of output. 

For observations from states other than $\rho$ and for the used range of thresholds we do not see any lines longer than $\Delta t$, hence it is impossible to estimate $K_2$. 
Moreover,  in Fig.~\ref{fig:502Lmax} we can see that the length of the longest line for the data (thick red line) and its surrogates (grey lines) do not differ.
We conclude that the origin of these observations is sufficiently well described by the assumption of linearly autocorrelated Gaussian noise. 
We can also try to soften our null hypothesis and use the shuffled surrogates.
It turns out, that all the observations\oldbtxt{with short lines only} are also perfectly consistent with this null hypothesis, thus here is no evidence of any dynamics at all (in Fig.~\ref{fig:502Lmax} the shuffled surrogates are plotted by cyan color).

\oldbtxt{Next group of observations are those, which} do contain longer diagonal lines (usually for higher $\epsilon$), but the values of $L_{\rm max}$ for data and \oldbtxt{IAAFT} surrogates do not differ notably. 
Within our sample of observations of IGR J17091-3624 only the observations 05-01 and 06-03 can be assigned to this subgroup, the example for 05-01 can be seen in Fig.~\ref{fig:501Lmax}.
However, comparing the data with the shuffled surrogates (cyan lines in Fig.~\ref{fig:501Lmax}), we find a striking difference, meaning \oldbtxt{that those observations cannot be described as temporally independent identically distributed random data.}
Hence in this case our analysis suggests that the data comes from a linear dynamical process (linearly autocorrelated Gaussian noise).

The \oldbtxt{rest of $\rho$} observations show stronger difference between the data and its surrogates, $L_{\rm max}^{\rm obs}$ is higher than $L_{\rm max}^{\rm surr}$ for all surrogates for wide range of $\epsilon$. This behaviour indicates non-linear features of the background dynamics. 


In order to quantify these results, we compute the estimate of $K_2$ for observations \oldbtxt{with enough long lines} and we will use this quantity as our discriminating quantity. The computation goes as follows.

\begin{figure}[t]
\includegraphics[width=\columnwidth]{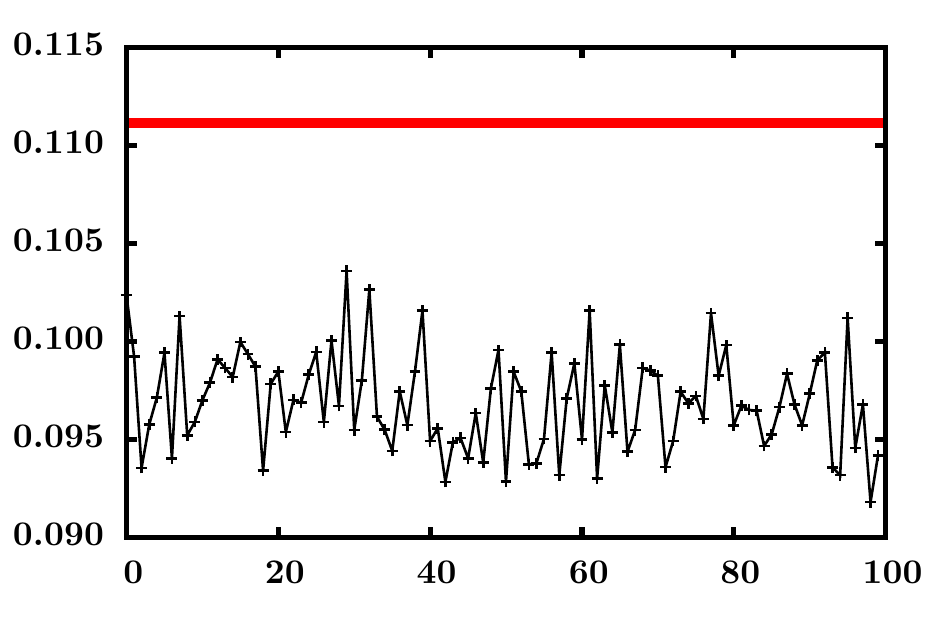}
\caption{ Comparison of recurrence rate for observation 06-02 (value is indicated as the horizontal thick red line) and its hundred surrogates (number of surrogate is on the $x$-axis) computed for the same recurrence parameters ($\epsilon=2.8, m=10,\Delta t = 14{\rm d} t$).
}
\label{fig:602RR_surr}
\end{figure}

For a chosen value of $\epsilon$ and the chosen embedding dimension and delay ($m=10$  and $\Delta t=7\jd{s}$) we compute the recurrence matrix and the cumulative histogram for the data and for \oldbtxt{its} surrogates. 
Usually  the recurrence rate for the surrogates is lower, which means that the shuffled order of points yields more distant points after the reconstruction of the phase space trajectory (see Fig.~\ref{fig:602RR_surr}), however for higher recurrence thresholds this difference is smeared. 
\oldbtxt{Afterwards, we estimate} $K_2$ from the slope of the cumulative histogram for the central part of the histogram which satisfies $l>\Delta t$, $p_c(\epsilon,l)>N_{\rm min}$ and $l<0.1N{\rm d}t$. 
We also require that there is at least 5 points of the cumulative histogram, which satisfy our criteria and could be used for the linear regression. 
\oldbtxt{The value of} $N_{\rm min}$ has to be chosen with respect to the length of the time series and the total number of lines. 
We usually choose $N_{\rm min}=100$. 
The example of cumulative histogram computed for the observation 06-02 with $\epsilon=2.8$ is given in Fig.~\ref{fig:602histogram}, where the used region of the histogram is marked by two vertical dashed lines.

\begin{figure}
\includegraphics[width=\columnwidth]{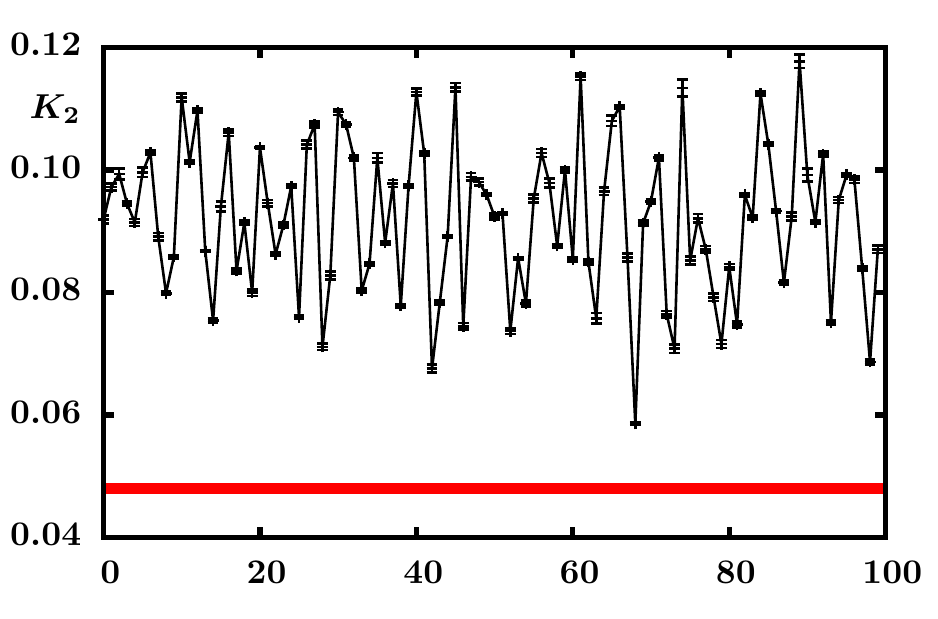}
\caption{ Comparison of estimates of $K2$ for observation 06-02 (indicated by horizontal thick red line) and its hundred surrogates (number of surrogate is on the $x$-axis) computed for the same recurrence parameters ($\epsilon=3.25, m=10,\Delta t = 14{\rm d} t$).
}
\label{fig:602K2_surr}
\end{figure}

\begin{figure}
\includegraphics[width=\columnwidth]{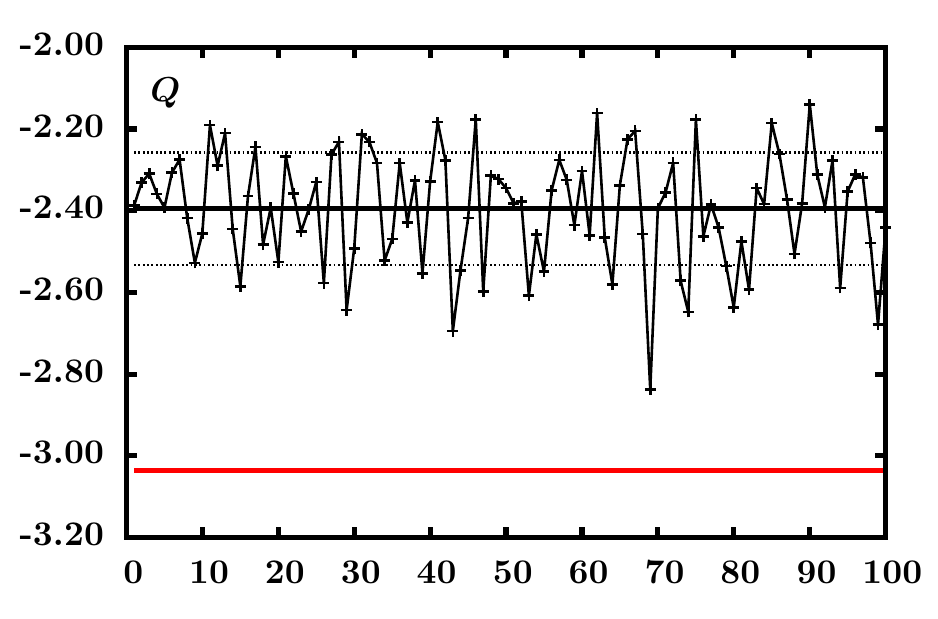}
\caption{ Comparison of $Q$ for observation 06-02 (indicated by horizontal thick red line) and its hundred surrogates (number of surrogate is on the $x$-axis) computed for the same recurrence parameters ($\epsilon=3.25, m=10,\Delta t = 14{\rm d} t$). The mean value $\bar{Q}$ and the standard deviation of the ensemble of surrogates is indicated by horizontal lines.}
\label{fig:602Q}
\end{figure}

We execute the same analysis with the same parameters on the set of surrogates \oldbtxt{and} we obtain set of $K_2^{\rm surr}$ values (see Fig.~\ref{fig:602K2_surr}).
According to \cite{Theiler199277} we define a significance of the obtained result as a weighted difference between the estimate of R\' enyi's entropy for the observed data $K_2^{\rm obs}$ and the mean of estimates of R\' enyi's entropy for ensembles of the surrogates $\bar{K}_2^{\rm surr}$.
However, as can be seen from relation \ref{cumul} the quantity $K_2$ is non-negative and behaves in exponential manner (for 10-times longer diagonal lines, $K_2$ drops down by one order). 
Therefore we first introduce the quantity $Q = \ln (K_2)$, which scales linearly, and then define the significance stemming from the estimate of $K_2$ \oldbtxt{as}
\begin{equation}
\mathcal{S}_{K_2} (\epsilon) = \frac{| Q^{\rm obs} (\epsilon) - \bar{Q}^{\rm surr}(\epsilon) |}{\sigma_{Q^{\rm surr}(\epsilon)}},
\end{equation}
where $\sigma_{Q^{\rm surr}}$ is the standard deviation of the set ${ Q^{\rm surr}_i}$ (see Fig.~\ref{fig:602Q}). 


When the observed data yields $K_2^{\rm obs}$, but the RP of its surrogates does not contain enough long lines, it is impossible to compute $Q^{\rm surr}$.
 Obviously this should contribute to the significance in the case, that $K_2^{\rm obs}$ is lower than $\bar{K}_2^{\rm surr}$. 
However, the values of $Q^{\rm surr}$ are not available. 
 \oldbtxt{We set the significance of the case}, when all $N^{\rm surr}$ surrogates have only short lines, to be $\mathcal{S}_{\rm sl}=3$.

On the other hand, if $K_2^{\rm obs}$ is higher than $\bar{K}_2^{\rm surr}$, then some of the surrogates have lines longer than the observed data.\oldbtxt{With that} the\oldbtxt{coincident} existence of surrogates, which do not have long lines at all, lessens the significance. 
This fact is incorporated in our procedure in the way that we add or subtract the two significances  $\mathcal{S}_{\rm sl}$ and  $\mathcal{S}_{K_2}$ according to the sign of the difference  $ Q^{\rm obs} (\epsilon) - \bar{Q}^{\rm surr}(\epsilon)$.


Finally, we obtain the total significance as the weighted difference of significances of these two cases, namely
\begin{equation}
\mathcal{S}(\epsilon) = \frac{N_{\rm sl}}{N^{\rm surr}} \mathcal{S}_{\rm sl} - {\rm sign}( Q^{\rm obs} (\epsilon) - \bar{Q}^{\rm surr}(\epsilon) ) \frac{N_{\mathcal{S}_K}}{N^{\rm surr}}  \mathcal{S}_{K_2}(\epsilon) , \label{significance}
\end{equation}
where $N_{\rm sl}$ is the number of surrogates, which have only short lines, and $N^{\rm surr}=N_{\mathcal{S}_K}+N_{\rm sl}$ is the total number of surrogates.
Our definition of significance yields negative values for some cases, e.g. when $K_2^{\rm obs} > \bar{K}_2^{\rm surr}$ and $N_{\rm sl}=0$. 
Since we expect, that the real non-linear dynamical system should contain more long lines than the artificial surrogates, only positive values of significance serve as the indication of non-linear dynamics. 
In fact, we use the value $\mathcal{S}=1.5$ as the limit, above which we consider the observation to be produced by non-linear dynamical system.
 
The last ambiguous step in the calculation of the significance is the choice of the recurrence threshold $\epsilon$. 
There are two \oldbtxt{possible ways how} to treat this issue. 
One way is to choose some recurrence rate for the observation (e.g. RR $\sim$ 10\%), choose the threshold appropriately and compute the significance. 
This approach is suitable for quick analysis, however it is not very accurate. 
In this paper we adopted the other way, which lies in finding the range of $\epsilon$, for which RR $\in (1\%,25\%)$, computing the significance for a linearly spaced set of $\epsilon$ in this interval (typically, we have used $N_\epsilon \sim 10 - 40$ values of $\epsilon$ in dependence on the number of points of the data set) and establishing the final decisive value as the average of the obtained significances $\bar{\mathcal{S}} = 1/N_\epsilon  \sum_\epsilon \mathcal{S}(\epsilon)$.
If for some $\epsilon$ the observed data set does not have enough long lines for the $K_2$ estimation, this value is not taken into account in the evaluation.

In  Table~\ref{TableIGR_result} \oldbtxt{we summarized} the length of the longest diagonal line (expressed as the number of points) for three different recurrence rates RR$\sim$5\%, 10\% and 20\%, the resulting averaged significance $\bar{\mathcal{S}}$ and the number of values of $\epsilon$ used for the averaging. 
All $\rho$ observations except of \oldbtxt{05-01 and 06-03 (recognized earlier as special group with respect to $L_{\rm max}$ behaviour)} have $\bar{\mathcal{S}}>1.5$.

In the same way we can also compute the significance $\bar{\mathcal{S}}_{\rm shf}$ for the other null hypothesis using the set of shuffled surrogates. 
This quantity serves as the indicator of departure from stochastic behaviour. 
Typically, we compute this quantity only for those observations, which have enough long lines but show non-significant result with the IAAFT surrogates.
This is because the data sets which have significant result for the non-linear dynamics should automatically have significant result for non-stochastic nature. 
However,  for the confirmation of this assumption in Section~\ref{sect:other_sources} we also provide several examples of this quantity for observations displaying the non-linear features.

\section{Testing the method with simulated time series}
\label{sect:poincare}

In general, our method can be applied to different kinds of time series, which are the product of any dynamical system, not only to X-ray lightcurves. For testing it is desirable to apply the method on time series, whose nature is known and which has been studied by other means. Because the dynamical system behind the observed lightcurves is a priori not known, we rather use numerical data.  \oldbtxt{We chose the numerical time series, which describe the motion of geodesic test particle in the field of static black hole surrounded by a massive thin disc. }

\oldbtxt{The details about the testing procedure will be given in separate paper \citep{chaos_proc}.  Here we only summarise the main results.}

Our analysis performed on the numerical time series revealed what the recurrence analysis provides for regular and chaotic trajectories of the complicated non-linear system, whose phase space consists of regular KAM tori nested in the chaotic layers. We have seen that even though governed by the same global set of equations, the evolution of the orbits in the regular and chaotic parts of the phase space can differ significantly, which is also reflected by the patterns in the recurrence plots. In agreement with the fact, that regular orbits in conservative systems have zero Lyapunov exponents, the estimate of $K_2$ is much lower for the regular trajectory than for its chaotic counterpart. 

Despite the expectation, that regular motion should not exhibit non-linear behaviour in our analysis, the significance for the regular orbit is quite high, however still almost twice lower than for the chaotic one. We argue that the reason for this is the way how the surrogates data are constructed, which means that they have exactly the same value distribution but they reproduce the spectrum only approximately, depending also on the available length of the data set. In case of the regular motion, very narrow peaks are in the spectrum and the error in reproducing such spectrum causes the very long diagonal lines to be broken. 


However for increasing strength of the added noise, the significance of the regular orbit drops down very quickly, while the chaotic orbit shows solid signs of non-linear behaviour even in the case, when the added noise has the same variance as the original data. Yet, even for high noise levels, $L_{\rm max}$ for the regular orbit is higher or similar as for the chaotic orbit.
Hence, observations which show non-significant result with the IAAFT surrogates, but very significant result for shuffled surrogates, and contain long diagonal lines compared to the time delay $\Delta t$, can possibly be explained as regular motion. Another possibility is, that such observations cover the part of the chaotic trajectory, which is called ``sticky orbit''. Because the observations cover only limited time interval, it is not possible to distinguish between these two cases.

\end{appendix}

\end{document}